\documentclass[onecolumn,12pt]{IEEEtran}
\usepackage{amsfonts,cite,graphicx,color}
\usepackage[cmex10]{amsmath}
\begin{document}

\title{The Effects of Limited Channel Knowledge on Cognitive Radio System Capacity}
\author{Peter~J.~Smith,~\IEEEmembership{Senior~Member,~IEEE},~Pawel~A.~Dmochowski,~\IEEEmembership{Senior~Member,~IEEE},~Himal~A.~Suraweera,~\IEEEmembership{Member,~IEEE},~and~Mansoor~Shafi,~\IEEEmembership{Fellow,~IEEE}%
\thanks{P. J. Smith is with the Department of Electrical and Computer Engineering, University of Canterbury, Christchurch, New Zealand (email: peter.smith@canterbury.ac.nz).}%
\thanks{P. A. Dmochowski is with the School of Engineering and Computer Science, Victoria University of Wellington, Wellington, New Zealand (email: pdmochowski@ieee.org).}%
\thanks{H. A. Suraweera is with the Singapore University of Technology and Design, Singapore (email: himalsuraweera@sutd.edu.sg).}%
\thanks{M.Shafi is with Telecom New Zealand, PO Box 293, Wellington, New Zealand (email: mansoor.shafi@telecom.co.nz).}%
\thanks{Parts of this work have been published at the \textit{IEEE Vehicular Technology Conference} (VTC2010-Fall), Ottawa, Canada, September 2010.}
%\thanks{Manuscript received December 31, 2008; revised January 1, 2009.}
}

\newcommand{\raisecaption}{\vspace{-0.0cm}}
\newcommand{\incgraphicswidth}{\includegraphics[width=0.95\linewidth]}
\newcommand{\ty}{\tilde{y}}
\newcommand{\Osp}{\Omega_\mathrm{sp}}
\newcommand{\Ops}{\Omega_\mathrm{ps}}
\newcommand{\Os}{\Omega_\mathrm{s}}
\newcommand{\Op}{\Omega_\mathrm{p}}
\newcommand{\Pp}{P_\mathrm{p}}
\newcommand{\Ps}{P_\mathrm{s}}
\newcommand{\Pt}{P_\mathrm{t}}
\newcommand{\Pm}{P_\mathrm{m}}
\newcommand{\ssp}{\sigma^2_\mathrm{p}}
\newcommand{\sss}{\sigma^2_\mathrm{s}}
\newcommand{\gT}{\gamma_\mathrm{T}}
\newcommand{\gamp}{\gamma_\mathrm{p}}
\newcommand{\gamI}{\gamma_\mathrm{I}}
\newcommand{\gp}{{g}_\mathrm{p}}
\newcommand{\gsp}{ {g}_\mathrm{sp}}
\newcommand{\gs}{{g}_\mathrm{s}}
\newcommand{\gps}{ {g}_\mathrm{ps}}
\newcommand{\hgp}{\hat{g}_\mathrm{p}}
\newcommand{\hgsp}{\hat{g}_\mathrm{sp}}
\newcommand{\hhp}{\hat{h}_\mathrm{p}}
\newcommand{\hhsp}{\hat{h}_\mathrm{sp}}
\newcommand{\tep}{\tilde{e}_\mathrm{p}}
\newcommand{\tesp}{\tilde{e}_\mathrm{sp}}

\maketitle

\begin{abstract}
We examine the impact of limited channel knowledge on the secondary user (SU) in a cognitive radio system. Under a minimum signal-to-interference-and-noise ratio (SINR) constraint for the primary user (PU) receiver, we determine the SU capacity under five channel knowledge scenarios. We derive analytical expressions for the capacity cumulative distribution functions and the probability of SU blocking as a function of allowable interference. We show that imperfect knowledge of the PU-PU channel gain by the SU-Tx often prohibits SU transmission or necessitates a high interference level at the PU. We also show that errored knowledge of the PU-PU channel is more beneficial than statistical channel knowledge  and imperfect knowledge of the SU-Tx to PU-Rx link has a limited impact on SU capacity.
\end{abstract}

\IEEEpeerreviewmaketitle

%%%%%%%%%%%%%%%%%%%%%%%%%%%%%%%%%%%%%%%%%%%%%%%%%
\section{Introduction}\label{Introduction}
The cognitive radio (CR) concept, introduced in \cite{Mitola2000}, refers to a smart radio which can sense the external electromagnetic environment and adapt its transmission parameters according to the current state of the environment~\cite{Haykin2005}. Secondary (or cognitive) users (SUs) can be designed to access parts of the primary user (PU) spectrum opportunistically or concurrently%
%for the information transmission
, provided that they cause minimal interference to the PUs in that band  \cite{Weiss2004, Wang2011, Shin2010}.

The CRs can protect the PU transmissions by a variety of control mechanisms. For example, the SU can regulate transmit power so that the interference at the PU receiver (PU-Rx) is below a well defined threshold. The limits on this received interference level can be imposed with an average and/or peak level constraint \cite{Ghasemi2007}. Another method of protecting the PU transmission is to consider a minimum value for its signal-to-interference noise ratio (SINR) beyond which further degradation is not accepted. Note that if the PU signal has a signal-to-noise ratio (SNR) below this level then the SU cannot transmit at all since the minimum SINR is unobtainable. With the SINR constraint, depending on the fading level of the PU transmitter (PU-Tx) to PU-Rx link, the conservatism inherent in the constant interference threshold constraint can be relaxed to some extent. The tolerable PU interference is no longer a constant and this can be to the benefit of the SU-Tx when the PU link is strong. The price of this relaxation is that some information about the PU-Tx to PU-Rx link must be available to the SU transmitter (SU-Tx).

A large body of work is now available on various aspects of CR systems, including fundamental information theoretic capacity limits and performance analysis, which often assumes perfect SU-Tx to PU-Rx channel state information (CSI) \cite{Jafar2007,Ghasemi2007,Musavian2009,Suraweera2008,Zhang2008,Kang2009,Wang2009}. In practice, there is expected to be limited (or no) collaboration between PU and SU systems. Hence, accurately estimating the SU-Tx to PU-Rx channels is a challenging task. An important question is the impact of the nature of channel knowledge on CR capacity. Several recent contributions have considered imperfect CSI \cite{Aissa2009,ShafiTVT2010,Alouini2011,AlouiniPIRMC2011,Tang2010,Popovski2011,Pei2011}. In \cite{Aissa2009}, mean and outage capacities along with optimum power allocation policies have been investigated for a CR system in a fading environment with imperfect CSI.

In order to enforce PU protection with imperfect CSI, probabilistic constraints are imposed in \cite{Aissa2009} and are also employed in this paper. The constraints in \cite{Aissa2009} guarantee that the interference power experienced by the PU receiver stays below a tolerable level within a prescribed probability. In our work, the probabilistic constraints apply to SINR. Assuming imperfect and quantized CSI for the SU-Tx to PU-Tx link, in \cite{ShafiTVT2010}, the mean capacity with peak interference power constraints was studied. In \cite{Alouini2011}, mean SU capacity is derived under average and peak transmit-power constraints and with respect to two different interference constraints: an interference outage constraint and a signal-to-interference outage constraint. To protect the PU under imperfect CSI, a new power control strategy is developed in \cite{Tang2010}. For a range of different assumptions about the available CSI at the SU-Tx, in \cite{Popovski2011}, achievable PU and SU rate regions were studied. The optimal robust transmitter design problem for a multiple-input single-output secure SU network under imperfect CSI was considered in \cite{Pei2011}. All of these studies have focused on the effect of imperfect knowledge of the SU-Tx to PU-Rx link, while ignoring the impact of imperfect CSI knowledge of other links, such as the PU-Tx to PU-Rx link at the SU-Tx. In \cite{Zhang2008}, it was demonstrated that obtaining the CSI of such links is highly beneficial to the SU capacity.

In \cite{AlouiniISWCS2010} and \cite{Alouini2011}, optimal power allocation and mean channel capacity is investigated for a secondary system under limited channel knowledge of the SU-Tx to the PU-Rx link. Both average and peak transmit power constraints are considered and two different interference constraints: an interference outage constraint and a signal-to-interference outage constraint. In contrast, \cite{AlouiniPIRMC2011} considers CSI imperfections on both the SU-Tx to the PU-Rx link and the SU-Tx to the SU-Rx link. Considering an average SU transmit power constraint and an instantaneous interference outage constraint, the authors have found an expression for the ergodic SU capacity.
 
The system considered in \cite{Dey2012} includes a single SU-Tx and SU-Rx pair sharing the same narrowband channel with $N$-multiple PUs. For this system, by considering various forms of imperfect CSI of the SU-Tx to the PU-Rx link at the secondary transmitter, the authors of \cite{Dey2012} have analyzed the mean SU capacity under an average SU transmit power constraint and  $N$ individual peak interference power constraints at each PU-Rx.

Some results in the case of multiple antenna deployments also exist in the literature. For example, in \cite{Duan2010}, the capacity of a spectrum sharing system with maximal ratio combining (MRC) diversity at the secondary receiver with imperfect CSI on the SU-Tx to the PU-Rx link is studied. Their results show that deployment of a multi-antenna array with MRC allows the secondary system to achieve a higher capacity as well as the opportunity to tolerate larger estimation errors.

This paper differs from the existing literature in several ways.  There are four channel links in a two user PU/SU channel to consider and each of them may or may not be perfectly known at the SU transmitter. Previous studies \cite{Aissa2009,ShafiTVT2010,Pei2011,Tang2010,Alouini2011} have only assumed imperfect knowledge of the SU-Tx to PU-Rx link. Thus, the impact of imperfect knowledge of the other links has not received a comprehensive treatment and remains unknown. Additionally, in previous work, the effect of the interference from the PU-Tx on SU capacity is ignored. Also, we employ the SINR at the PU-Rx to impose probabilistic constraints to protect the PU-Rx, while prior works, with the exception of \cite{Alouini2011}, have considered an interference outage constraint. Finally, we consider several cases where the imperfect CSI manifests itself in the form of statistical channel knowledge (i.e., knowledge of the mean channel gains). Such a form of imperfect CSI is attractive from a practical stand point, since obtaining accurate knowledge is almost impossible for some links, such as the PU-Tx to PU-Rx link. Moreover, the mean value does not impose a large system burden as it only requires infrequent updates. Note that the inclusion of PU-Tx to SU-Rx interference  and probabilistic constraints enables a rigorous evaluation of the benefits of various types of CSI. However, it also increases the analytical complexity. Hence, in order to make progress on the key issue of assessing the impact of CSI we focus on the simple case of a two user PU/SU channel.

In this paper, we study the SU performance under various scenarios for the CSI available at the transmitter with the aim of evaluating the relative importance of the different links in our model. We consider knowledge of the mean and errored channel gains as types of imperfect CSI and the baseline case of perfect CSI. All three situations are relevant with perfect CSI providing a benchmark and imperfect CSI representing some level of SU-PU cooperation. The case where only the mean channel gains are available corresponds to very low rate feedback of the mean values from the receivers to the SU-Tx. The channel needs to be updated only when the positions of the terminals are changed. %At the PU-Rx we consider a SINR constraint which is more flexible for SU operation. 
%%%%%%%%%%%%%%%%%%%%%%%%%%%%%%%
%Studies pertaining to a primary SINR constraint, as compared to the popular interference temperature constraint, are limited in the literature largely due to the %mathematical complexity required in an analysis. 
%Hence, in this paper we focus on the effects of limited CSI on SU operation with an SINR constraint at the PU-Rx. It is difficult to compare systems fairly when they have %different levels of CSI. 
In all cases where the channels are not exactly known, we are able to compare performance using a single probabilistic SINR constraint. This provides novel and fair comparisons and allows the true importance of various CSI levels to be identified. In particular, we establish the following key observations and results:
\begin{itemize}
	\item In four of the five scenarios considered, we derive analytical expressions for the cumulative distribution function (cdf) of the SU SINR and use it to evaluate the SU capacity cdf. 
	\item For all scenarios, we derive the probability of SU blocking as a function of the permissible interference at the PU-Rx.
	\item By evaluating our results for a range of system parameters, we demonstrate the importance of accurate knowledge of the PU-Tx to PU-Rx link at the SU-Tx. We show that under demanding operating conditions, this information is required for SU transmission, in contrast to the full knowledge of other links.
	\item We demonstrate the very high sensitivity of SU performance to the error in the estimation of the PU-Tx to PU-Rx and SU-Tx to PU-Rx links.
  \item We show that errored knowledge of the PU-Tx to PU-Rx  link and  SU-Tx to PU-Rx link (if available) is better for SU capacity than a knowledge of the mean link gains and, of the two, the former has more impact on SU capacity.
	\item By considering a single probabilistic SINR constraint, a unified framework is presented which enables fair comparisons between different types of channel knowledge. Hence, intuitive results such as the importance of the PU-Tx to PU-Rx link, can be verified and isolated results from the literature can be confirmed in a rigorous manner. 
\end{itemize}

%In this paper we consider the effects of imperfect CSI and the inclusion of the interference to the SU due the PU transmission. We derive expressions for the SU capacity cdfs as well as the SU blocking probability. In doing so, we demonstrate the importance of accurate knowledge of the PU-Tx to PU-Rx link to the SU performance.

The rest of the paper is organized as follows. Section II introduces the system model. In Section III, we investigate the mean SU capacity and SU blocking probability under different channel knowledge scenarios. In Section IV, numerical results supported by simulations are presented and discussed. Finally, we conclude in Section V.

%%%%%%%%%%%%%%%%%%%%%%%%%%%%%%%%%%%%%%%%%%%%%%%%%%%%%%%%%%
\section{System Model}
\begin{figure}[e]
	\centering \includegraphics[width=0.7\columnwidth]{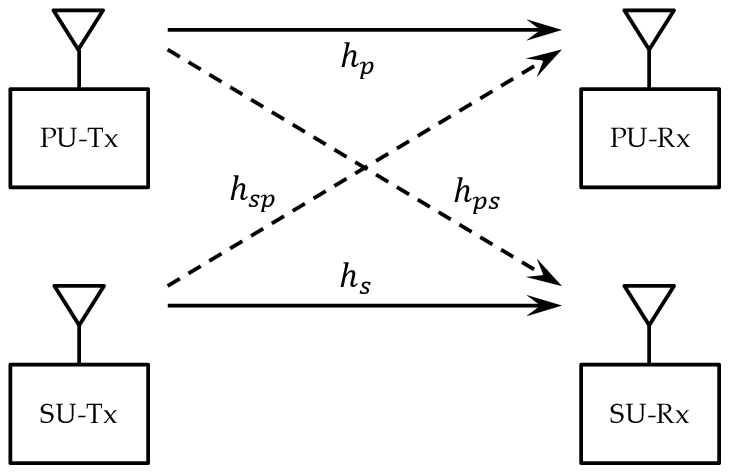}
	\raisecaption\caption{System Diagram.}
	\label{fig:SystemDiagram}
\end{figure}

%
%As shown in Fig. \ref{fig:SystemDiagram}, 
Consider a CR system (shown in Fig. \ref{fig:SystemDiagram}) with the SU-Tx and PU-Tx transmitting simultaneously to their respective receivers. Independent point-to-point flat Rayleigh fading channels are assumed for all links. Let $\gp=|h_\mathrm{p}|^2$, $\gs =|h_\mathrm{s}|^2$, $\gps=|h_\mathrm{ps}|^2$ and $\gsp=|h_\mathrm{sp}|^2$ denote the instantaneous channel gains of the PU-Tx to PU-Rx, SU-Tx to SU-Rx, PU-Tx to SU-Rx and SU-Tx to PU-Rx links, respectively. Denote the exponentially distributed probability density functions (pdfs) of the random variables (RVs) $\gp$, $\gs $, $\gps$ and $\gsp$ by $f_{\gp}(x)$, $f_{\gs }(x)$, $f_{\gps}(x)$ and $f_{\gsp}(x)$, respectively, governed by their corresponding parameters, $\Op =\mathbb{E}(\gp), \Os =\mathbb{E}(\gs )$, $\Ops =\mathbb{E}(\gps)$ and $\Osp =\mathbb{E}(\gsp)$, where $\mathbb{E}(\cdot)$ denotes the expectation operator.

As described further in this Section, the SU transmission under the SINR constraint is governed solely by the state of the $\gp$ and $\gsp$ links%
\footnote{The channel gains $\gs$ and $\gps$ have an impact on achievable SU capacity, however the level of their knowledge by the SU-Tx does not impact the transmit power $\Ps$.}%
. Thus, in this paper we consider the following five scenarios for the knowledge of $\gp$ and $\gsp$ by the SU-Tx.

\emph{Scenario 1}: The PU-Tx to PU-Rx channel, $\gp$, and the SU-Tx to PU-Rx channel, $\gsp$, are perfectly known. This clearly unrealistic scenario serves as a benchmark for comparison of the other cases. 

\emph{Scenario 2}: The PU-Tx to PU-Rx channel, $\gp$, is perfectly known while only the mean $\Osp $ of the channel between the SU-Tx and the PU-Rx is known. We consider this scenario to reflect the fact that while the PU is more likely to estimate and feed back the full CSI of its own communication link, it should not be tasked to do so for the SU-Tx to PU-Rx channel. Instead, only statistical information about $\gsp$ is relayed back to SU-Tx.

\emph{Scenario 3}: The mean, $\Op$, and the exact channel gain, $\gsp$, are known. In contrast to \emph{Scenario 2}, this case is considered mainly for completeness. 

\emph{Scenario 4}: Only the means, $\Op$ and $\Osp$, are known. This scenario arises when only statistical information about the channels is available to the SU-Tx as a result of limited feedback resources. 

\emph{Scenario 5}: Estimates of the PU-Tx to PU-Rx and SU-Tx to PU-Rx channels are available. This scenario arises as a result of channel estimation errors, as well as feedback quantisation and delay. 

%In all cases, where possible, 
Where possible, %
we impose a constraint, $\gT $, on the PU-Rx SINR, denoted by $\gamp$. Hence,
\begin{align}\label{eq:SINRconstraint}
	\gamp  = \frac{\Pp \gp}{\Ps \gsp + \ssp },\  \textrm{and}\ \gamp \geq \gT ,
\end{align}
where $\gT $ is a pre-defined SINR threshold, $\Pp$ (assumed constant and known to the SU-Tx) and $\Ps$ are the PU and SU transmit powers, respectively, and $\ssp$ is the additive white Gaussian noise (AWGN) variance at the PU-Rx. In the event that the PU-Rx SNR lies in the region, $\Pp \gp/\ssp <\gT $, the constraint in \eqref{eq:SINRconstraint} cannot be satisfied, and thus, the SU transmit power is zero and thus there is no SU interferences to the PU. If the PU SNR is above the SINR threshold $\gT$, the SU-Tx will adapt its transmit power to a maximum level satisfying \eqref{eq:SINRconstraint} as determined under the five scenarios. This adaptation does not consider the SU link $\gs$. We also impose a maximum SU transmit power constraint, $\Pm $, which arises in practice, for example, due to power amplifier nonlinearities. Thus, in \textit{Scenario} 1, where the SU-Tx knows $\gp$, the SU transmit power is given by
%
%\begin{align}\label{eq:Pt}
	%\Pt =	\begin{cases}
					%0 & \frac{\Pp  \gp}{\gT }<\ssp \\
					%\min\left(\Ps, \Pm  \right) & \textrm{otherwise},
				%\end{cases}
%\end{align}
%%
\begin{eqnarray} % more compact
\Pt &=& \left\{
\begin{array}{ll} 0 & \frac{\Pp  \gp}{\gT }<\ssp \\
								\min\left(\Ps, \Pm  \right) & \textrm{otherwise},
\end{array}\right.
\label{eq:Pt}
\end{eqnarray}
where $\Ps$ is obtained from \eqref{eq:SINRconstraint} by solving $\gT =\gamp$. 
%As will be shown in Section \ref{SUCapacity}, the $\Pt =0$ condition for \textit{Scenarios} 3-5 will depend on the choice of system parameters and thus \eqref{eq:Pt} will be modified accordingly.
%Furthermore, we note that
Furthermore, the constraints described above can only be guaranteed if the SU-Tx has perfect knowledge of the links $\gp$ and $\gsp$, i.e., under \textit{Scenario} 1. In analysing \textit{Scenarios} 2-5, we use probabilistic constraints. Hence, we require the SINR constraint to hold with an acceptably high probability, $1-\alpha$, where $\alpha$ is small.  These are described in Sections \ref{Scenario3} - \ref{Scenario5}. %Sections \ref{Scenario3}, \ref{Scenario4} and \ref{Scenario5}.

In analysing the SU capacity, we first consider the SINR at the SU-Rx, denoted by $\gamI$, where
\begin{align}\label{eq:gammaI}
	\gamI = \frac{\Pt  \gs }{\Pp  \gps + \sss},
\end{align}
and $\sss$ is the AWGN variance at the SU-Rx. We denote the pdf and cdf of $\gamI$  by $f_{\gamI}(x)$ and $F_{\gamI}(x)$, respectively. The instantaneous SU capacity is given by
\begin{align}\label{eq:instantC}
	C = \log_2\left(1+\gamI\right),
\end{align}
where the mean, $\bar{C}$, can be derived using $f_{\gamI}(x)$ as
\begin{align}\label{eq:meanC}
	\bar{C}=\mathbb{E}(C)= \int ^{\infty}_{0}\log_2\left(1+x\right)f_{\gamI}(x) \ dx.
\end{align}

The cdf of $C$ can be obtained from $F_{\gamI}(x)$ by noting that
\begin{align}\label{eq:FCgenI}
	F_C(y)=\textrm{Pr}(\gamI<2^y-1)=F_{\gamI}(\tilde{y}),
\end{align}
where $\textrm{Pr}(\cdot)$ denotes probability and $\tilde{y}=2^y-1$. Using \eqref{eq:gammaI}, we can express \eqref{eq:FCgenI} as
\begin{align}\label{eq:FgIgen}
	F_{\gamI}(\tilde{y})&=\mathbb{E}_{\gps}\left\{ \textrm{Pr} \left( \Pt  \gs < \tilde{y} (\sss+\Pp  \gps) \right) \Biggl| \gps \right\} \\ \nonumber
	&=\int_{0}^{\infty}F_{\gamma}\left( \tilde{y}(\sss+\Pp  v)\right) \frac{e^{-v/\Ops }}{\Ops } \ dv,
\end{align}
where we have defined $\gamma=\Pt \gs $ with a cdf $F_{\gamma}(x)$. In what follows, we derive expressions for $F_{\gamma}(x)$ which, using \eqref{eq:FCgenI} and \eqref{eq:FgIgen}, allows us to compute the capacity cdf.

We parameterize the main system variables by two key parameters. %
%The main system variables can be parameterized using two key parameters. 
The first, $c_1$, defined by
\begin{align}\label{eq:c1}
	c_1=\frac{\Osp }{\Os },
\end{align}
represents the ratio of interference at the PU-Rx to the desired channel strength for the SU.  The second, $c_2$, is given by
\begin{align}\label{eq:c2}
	c_2=\frac{\gT }{\Pp \Op /\ssp },
\end{align}
which is the ratio of the minimum target SINR to the mean SNR at the PU-Rx. Hence, increasing $c_2$ corresponds to reducing the allowable interference, with the case of $c_2=1$ corresponding to zero average allowable interference.

%%%%%%%%%%%%%%%%%%%%%%%%%%%%%%%%%%%%%%%%%%%%%%%%
\section{SU Capacity}\label{SUCapacity}
The capacity mean in \eqref{eq:meanC} and the cdf in \eqref{eq:FgIgen} require a knowledge of the distributions of $\gamma=\Pt \gs$ and $\gamI$. Hence, in this section we derive the cdfs for $\gamma$ and $\gamI$ for \textit{Scenarios} 1-4. For \textit{Scenario} 5, an alternative approach is required (see Section \ref{Scenario5}).

%%%%%%
\subsection{Scenario 1} \label{Scenario1}

When the SU has full knowledge of $\gp$ and $\gsp$, $\Ps$ can be obtained directly from \eqref{eq:SINRconstraint}, giving
\begin{align}\label{eq:PsScen1}
	\Ps=	\frac{\frac{\Pp  \gp}{\gT }-\ssp }{\gsp}.
\end{align}
We note that while we ignore the $\Pt =0$ case in \eqref{eq:Pt}, the following derivation is valid since $\textrm{Pr}(\gamma>0)=0$ for $\Pt  \leq 0$. In finding $F_{\gamma}(x)$, we solve for the complementary cdf given by
\begin{align}\label{eq:PrgScen1a}
	\textrm{Pr}(\gamma>x)&=\textrm{Pr}\left(\gs  \min(\Pm ,\Ps)>x\right)  \nonumber \\
			&=\textrm{Pr}\left(\Pm  \gs  > x, \left(\frac{\Pp  \gp}{\gT }-\ssp \right)
						\frac{\gs }{\gsp}>x\right) \nonumber \\
			&=\textrm{Pr}\left(\gs  > \frac{x}{\Pm }, \left(\frac{\Pp  \gp}{\gT }-\ssp \right)
						\gs>x\gsp\right).
\end{align}
Noting that $\gp$ is an exponentially distributed RV, we can rewrite \eqref{eq:PrgScen1a} as
\begin{align}\label{eq:PrgScen1b}
	\textrm{Pr}(\gamma>x)=\int^{\infty}_{0}\int^{\infty}_{\frac{x}{\Pm }}
						e^{-\frac{\gT }{\Pp \Op }\left(\frac{x v}{u}+\ssp \right)}
						f_{\gs }(u)f_{\gsp}(v) \ du \ dv.
\end{align}
Substituting for $f_{\gs }(u)$ and $f_{\gsp}(v)$, and changing the order of integration one obtains
\begin{align}\label{eq:PrgScen1c}
	\textrm{Pr}(\gamma>x) %\\ \nonumber
		&=\frac{e^{-\frac{\gT \ssp }{\Pp \Op }}}{\Osp \Os }
			\int^{\infty}_{\frac{x}{\Pm }} e^{-\frac{u}{\Os }}
			\int^{\infty}_{0}e^{-\left(\frac{\gT  x}{\Pp  \Op  u}+\frac{1}{\Osp }\right)v} \ dv \ du \\ \nonumber
		&=\frac{e^{-\frac{\gT  \ssp }{\Pp \Op }}}
			{\Osp \Os }\int^{\infty}_{\frac{x}{\Pm }}\frac{e^{-\frac{u}{\Os }}}
			{\frac{\gT  x}{\Pp  \Op  u}+\frac{1}{\Osp }} \ du.
\end{align}
After simplifying \eqref{eq:PrgScen1c}, the cdf $F_{\gamma}(x)=1-\textrm{Pr}(\gamma>x)$ can be shown to be \cite[Eq. (3.351.2)]{Gradshteyn2007}
\begin{align}\label{eq:FgScen1}
		F_{\gamma}(x)&=1-e^{-\frac{\gT \ssp }{\Pp \Op }}
				\Biggl[e^{-\frac{x}{\Pm \Os }}-\frac{\Osp  \gT   x }{\Pp  \Op  \Os }
						e^{\frac{\Osp  \gT  x}{\Pp  \Op  \Os }} %\\ \nonumber	& \times
				 \Gamma\left(0,\frac{\Osp  \gT   x}{\Pp  \Op  \Os }
					+ \frac{x}{\Pm \Os }\right)\Biggr],
\end{align}
where $\Gamma(\cdot,\cdot)$ is the upper incomplete gamma function. %
%In order to derive $F_{\gamma_I}(\tilde{y})$, we substitute \eqref{eq:FgScen1} into \eqref{eq:FgIgen}, resulting in
Substituting \eqref{eq:FgScen1} into \eqref{eq:FgIgen} results in
\begin{align}\label{eq:FgIScen1}
		F_{\gamma_I}(\ty)&=1-\frac{\Pm \Os  e^{-\left(\frac{\gT \ssp }{\Pp \Op }+\frac{\ty\sss}{\Pm \Os } \right)}}
															{\Pm \Os +\ty \Pp \Ops } +									
										\frac{\Osp \gT \ty}{\Ops\Op\Os \Pp }
										\exp\left\{  \frac{\Osp \gT \ssp}{\Op \Os \Pp } \left(\ty -\frac{\Os}{\Osp}\right) \right\} \\ \nonumber
										& \times \int_{0}^{\infty} \left(\ssp+\Pp v \right) \exp \left\{\left(  \frac{\Osp \gT }{\Op \Os} \ty -\frac{1}{\Ops} \right)v\right\}
										\Gamma\left( 0, \frac{\Osp \gT \Pm +\Pp \Op } {\Pp  \Pm  \Op\Os} ( \ssp+\Pp  v )\ty \right) \ dv.
\end{align}
To the best of the authors' knowledge, the integral in \eqref{eq:FgIScen1} does not have a closed-form solution. In Section \ref{Simulations}, the capacity cdf results are obtained by numerical integration.

In order to obtain the expression for mean capacity, we can derive the pdf, $f_{\gamma}(x)$, by differentiating \eqref{eq:FgScen1} with respect to $\ty$. Alternatively, using \eqref{eq:FgIgen} we have
\begin{align}\label{eq:fgIgen}
	f_{\gamI}(\ty) = \int_{0}^{\infty}(\ssp+\Pp  v) f_{\gamma}(\ty(\ssp+\Pp  v)) \frac{e^{-v/\Ops}}{\Ops} \ dv,
\end{align}
where $f_{\gamma}(x)$ was computed in \cite{Dmochowski2010} as,
\begin{align}\label{eq:fgScen1}
	f_{\gamma}(x)&=e^{-\frac{\gT \ssp }{\Pp \Op }}\Biggl[\left(\frac{1}{\Pm  \Os }-\frac{\Osp \gT }{\Pp  \Op  \Os }\right)e^{-\frac{x}{\Pm \Os }}\\\nonumber
&+ e^{\frac{\Osp  \gT   x}{\Pp  \Op  \Os }}\left(\frac{(\Osp  \gT )^2 x}{(\Pp \Op \Os )^2}+\frac{\Osp \gT }{\Pp  \Op  \Os }\right) %\\\nonumber & \times
 \Gamma\left(0,\frac{\Osp  \gT   x}{\Pp  \Op  \Os } + \frac{x}{\Pm \Os }\right)\Biggr].
\end{align}
The expression resulting from substituting \eqref{eq:fgScen1} into \eqref{eq:fgIgen} cannot be written in closed-form. Thus, the mean capacity, $\bar{C}$, must be calculated numerically by substituting \eqref{eq:fgScen1} into \eqref{eq:fgIgen} and \eqref{eq:meanC}.
%
%%%%%%%%%%%%%%%%%%%%%%%
\subsection{Scenario 2}\label{Scenario2}
In \textit{Scenarios} 2-5, where exact channel knowledge is unavailable, the SU cannot guarantee that \eqref{eq:SINRconstraint} is satisfied since the values of $\gp $ and $\gsp$ are uncertain. Thus, we constrain the SU to satisfy \eqref{eq:SINRconstraint} with an acceptably high probability, $1-\alpha$, where $\alpha$ is usually small.

Hence, for \textit{Scenario} 2, where the SU knows only the mean, $\Osp $, of $\gsp$, we consider the probability of satisfying the SINR constraint with a probability of $1-\alpha$. This gives
\begin{align}\label{eq:PrScen2a}
	\textrm{Pr}&\left(\frac{\Pp  \gp}{\Ps \gsp + \ssp } \geq \gT  \Biggl|\gp, \Osp \right)=1-\alpha,
\end{align}
which can be rewritten as
\begin{align}\label{eq:PrScen2b}
	\textrm{Pr}\left(\gsp\leq \frac{\Pp  \gp-\gT  \ssp }{\Ps \gT } \Biggl|\gp, \Osp  \right)=1-\alpha.
\end{align}
Since $\gsp$ is exponential, from \eqref{eq:PrScen2b}, we can derive the expression for the transmit power, $\Ps$, as
\begin{align}\label{eq:PsScen2}
	\Ps=-\frac{\Pp  \gp-\gT \ssp }{\ln(\alpha)\gT  \Osp }.
\end{align}
Using \eqref{eq:PsScen2}, the complementary cdf of $\gamma$ is derived as follows:
\begin{align}\label{eq:PrgScen2aa}
		\textrm{Pr} \left(\gamma>x\right)
		&= 	\textrm{Pr}\left(\Pm  \gs  > x, \Ps \gs  > x \right)  \nonumber\\
		&= 	\mathbb{E}\left[\textrm{Pr}\left(\Pm  \gs  > x, \Ps \gs  > x \Bigl|\gp \right)\right],
		%\\ \nonumber
		%&= \begin{cases}	\textrm{Pr}\left(\gs  > \frac{x}{\Pm } \right) & \Pm  < \Ps  \\
				%\textrm{Pr}\left(\gs  > \frac{x}{\Ps}  \right) & \Pm  > \Ps,
				%\end{cases}
\end{align}
where the conditional probability in \eqref{eq:PrgScen2aa} is given by
\begin{align}\label{eq:PrgScen2a}
		\textrm{Pr}\left(\Pm  \gs  > x, \Ps \gs  > x \Bigl|\gp \right)
		= \begin{cases}	\textrm{Pr}\left(\gs  > \frac{x}{\Pm } \right) & \Pm  \leq \Ps  \\
				\textrm{Pr}\left(\gs  > \frac{x}{\Ps}  \right) & \Pm  > \Ps.
				\end{cases}
\end{align}
Hence, we have
\begin{align}\label{eq:PrgScen2b}
		\textrm{Pr} \left(\gamma>x\right)
		& = 	\int_{\psi_0}^{\psi}\textrm{Pr}\left(\gs  > \frac{x}{\Ps} \right) f_{\gp}(y) \ dy %\\ \nonumber &
		+ \int_{\psi}^{\infty}\textrm{Pr}\left(\gs  > \frac{x}{\Pm } \right) f_{\gp}(y) \ dy,
\end{align}
where $\psi_0=\frac{\gT  \ssp }{\Pp }$ and $\psi=\frac{\gT (\ssp -\Pm \ln(\alpha)\Osp )}{\Pp }$. The lower integration limit in the first term of \eqref{eq:PrgScen2b} takes into account the $\Pt =0$ condition in \eqref{eq:Pt}. After some manipulation, we can simplify \eqref{eq:PrgScen2b} to obtain $F_\gamma(x)=1-\textrm{Pr} \left(\gamma>x\right)$ as,
\begin{align}%\label{eq:FgScen2}
		F_{\gamma}(x)&=1-\int_{\psi_0}^{\psi} \textrm{Pr}\left(\gs  > \frac{x}{\Pm }\right)f_{\gp}(y) \ dy %\\ \nonumber &
									-\int^{\infty}_\psi \textrm{Pr}
									\left(\gs > -\frac{\ln(\alpha)\gT  \Osp x}{\Pp  y-\gT \ssp }\right)f_{\gp}(y) \ dy \label{eq:FgScen2} \\
%\end{align}
%%
%which results in
%%
%\begin{align}\label{eq:FgScen2b}
	%F_{\gamma}(x)	
	& =1-\exp\left\{ -\frac{x}{\Pm \Os }
										-\frac{\psi}{\Op }\right\} %\\ \nonumber &
								 -\frac{1}{\Op }\int_{\psi_0}^{\psi}
										e^{-\frac{\ln(\alpha)\gT  \Osp x}
										{(\gT \ssp -\Pp  y)\Os }}e^{-\frac{y}{\Op }} \ dy. \label{eq:FgScen2b}
\end{align}
Once again, there exists no closed-form solution to the integral in \eqref{eq:FgScen2b}. Following the same approach as in \textit{Scenario 1}, we use \eqref{eq:FgScen2b} and \eqref{eq:FgIgen} to find $F_{\gamI}(\ty)$ (and thus the capacity cdf from \eqref{eq:FCgenI}). After some manipulation, we obtain
\begin{align}\label{eq:FgIScen2}
	F_{\gamI}(\ty)	&=1-\frac{\Pm \Os e^{-\left( \frac{\ty\ssp}{\Pm \Os}+\frac{\psi}{\Op}\right)}}{\Ops \Pp  \ty +\Pm \Os} \\ \nonumber
									&+\frac{1}{\Os}\int_{\psi_0}^{\psi}
									e^{-\left( \frac{\Osp\gT\ssp\ln\alpha\ty}{\gT \ssp \Os-\Pp \Os z} +\frac{z}{\Os}\right)}
									\frac{\gT \ssp \Os-\Pp \Os z}{\gT \ssp \Os+\Osp \gT \Pp \Ops \ln(\alpha)\ty-\Pp \Os z} \ dz.
\end{align}
Here, again, the capacity cdf is obtained using \eqref{eq:FCgenI} and numerically integrating \eqref{eq:FgIScen2}.

To compute the SU mean capacity, we differentiate \eqref{eq:FgIScen2} with respect to $\ty$ to find the pdf
\begin{align}\label{eq:fgScen2}
	f_{\gamI}(x)	&= -\ssp e^{-\left( \frac{x \ssp}{\Pm \Os}+\frac{\psi}{\Op}\right)}
									+\frac{\Osp\gT\ssp\ln(\alpha)}{\Os} \int_{\psi_0}^{\psi}
									e^{-\left( \frac{\Osp\gT\ssp\ln(\alpha) x}{\gT \ssp \Os -\Pp \Os z} +\frac{z}{\Os}\right)} \\ \nonumber
									& \times \left(\frac{( \gT \ssp \Os  - \Pp  \Os z  ) (\Osp \gT \Pp \Ops \ln\alpha -1  ) +\Osp \gT \Pp \Ops \ln(\alpha) x }{(\gT \ssp \Os +\Osp \gT \Pp \Ops \ln(\alpha) x-\Pp \Os z)^2}\right) \ dz.
\end{align}
The mean capacity is then computed by substituting \eqref{eq:fgScen2} into \eqref{eq:meanC} and numerically integrating.

%%%%%%%%%%%%%%%%%%%%%%%
\subsection{Scenario 3}\label{Scenario3}
In \textit{Scenario} 3, where the SU has exact knowledge of $\gsp$ and knows only the mean $\Op $, we once again satisfy the SINR constraint with a probability of $1-\alpha$. Hence,
\begin{align}\label{eq:PrScen3a}
	\textrm{Pr}\left(\frac{\Pp  \gp}{\Ps \gsp + \ssp }
							\geq \gT  \Biggl| \Op , \gsp\right)=1-\alpha,
\end{align}
which gives
\begin{align}\label{eq:PrScen3b}
	\textrm{Pr}\left( \gp \geq \frac{\gT (\Ps \gsp+\ssp )}{\Pp } \Biggl| \Op , \gsp\right)=1-\alpha.
\end{align}
Following the same approach as for \textit{Scenario} 2 %in Sec. \ref{Scenario2}, noting that $\gp$ is an exponential random variable
, one can show that
\begin{align}\label{eq:PsScen3}
	\Ps=-\left(\frac{\ln(1-\alpha)\Pp \Op }{\gT }+\ssp \right)\frac{1}{\gsp}.
\end{align}
From \eqref{eq:PsScen3}, the SU SINR cdf $F_{\gamI}(\ty)$ and pdf $f_{\gamI}(\ty)$ are derived in Appendix \ref{appScen3}. The cdf is expressed in terms of simple functions of $\ty$ as
\begin{align}\label{eq:FgIScen3compMain}
		F_{\gamma_I}(\tilde{y}) & = 1-s(\ty) -h(\ty) \textrm{E}_1(r(\ty)),
\end{align}
where $s(\ty)$, $h(\ty)$ and $r(\ty)$ are given in Appendix \ref{appScen3} by \eqref{eq:shr}. Similarly, the pdf is given by
\begin{align}\label{eq:fgIScen3compMain}
		f_{\gamma_I}(\tilde{y}) & = -s'(\ty)-h'(\ty) \textrm{E}_1(r(\ty))+h(\ty)r'(\ty)\frac{e^{-r(\ty)}}{r(\ty)},
\end{align}
where $s'(\ty)$, $h'(\ty)$ and $r'(\ty)$ are given in Appendix \ref{appScen3} by \eqref{eq:dershr}.

%%%%%%%%%%%%%%%%%%%%%%%
\subsection{Scenario 4}\label{Scenario4}
Consider the scenario where the SU-Tx has knowledge of only the mean values of $\gp$ and $\gsp$. Here, we have
\begin{align}\label{eq:PrScen4a}
	\textrm{Pr}\left(\frac{\Pp \gp}{\Ps \gsp+\ssp } \geq \gT  \biggl| \Op ,\Osp \right)=1-\alpha.
\end{align}
Using conditioning, \eqref{eq:PrScen4a} can be given as
\begin{align}\label{eq:PrScen4b}
	\mathbb{E}\left[\textrm{Pr}\left(\Pp \gp\geq \gT \left(\Ps \gsp+\ssp \right) \biggl| \gsp\right)\right]=1-\alpha,
\end{align}
which after some manipulation gives the transmit power, $\Ps$,  as
\begin{align}\label{eq:PsScen4}
	\Ps=\frac{\Pp \Op }{\gT \Osp }\left( \frac{e^{-\frac{\gT \ssp }{\Pp \Op }}}{1-\alpha}-1  \right).
\end{align}
Here, $\Ps$ and $\Pt$ are deterministic, depending on the system parameters. The latter is given by
%%
%\begin{align}\label{eq:PtScen4}
	%\Pt =	\begin{cases}
				%0 & \Ps < 0 \\
				%\Ps & 0<\Ps<\Pm  \\
				%\Pm & \Ps>\Pm.
	%\end{cases}
%\end{align}
%
\begin{eqnarray} %more compact
\Pt &=& \left\{
\begin{array}{ll} 0 & \Ps < 0\\
								\Ps & 0<\Ps<\Pm  \\
								\Pm & \Ps>\Pm.
\end{array}\right.
\label{eq:PtScen4}
\end{eqnarray}
Hence, the cdf of $\gamma$ is given by
\begin{align}\label{eq:CapacityCDF4a}
	F_{\gamma}(x)=1-e^{-\frac{x}{\Pt \Os }},
\end{align}
which, after substituting into \eqref{eq:FgIgen} and \eqref{eq:FCgenI}, results in the capacity cdf
\begin{align}\label{eq:FCintScen4}
	F_C(y)=1-\frac{\Pt \Os }{\tilde{y}\Pp \Ops +\Pt \Os }e^{-\frac{\tilde{y}\sss }{\Pt \Os }}.
\end{align}
In order to compute the mean capacity, $\bar{C}$, we note that $F_{\gamma_I}(x)$ can be trivially derived from \eqref{eq:FCintScen4} and \eqref{eq:FgIgen}. %Upon
Differentiating to obtain $f_{\gamma_I}(x)$ and substituting into \eqref{eq:meanC}, one obtains %can show that the mean capacity is given by
\begin{align}
\label{eq:meanCScen4}
	\bar{C}=\frac{1}{\ln(2)}\int_{1}^{\infty}
	\left(	\frac{\sss  }{\Pp \Ops t+\Pt \Os }
			+		\frac{\Pt \Pp \Os \Ops  }{\left(\Pp \Ops t+\Pt \Os \right)^2}
				\right)				\ln(t)	e^{\frac{-t\sss }{\Pt \Os }} \ dt,
\end{align}
where we have used the change of variable, $t=1+x$.

%%%%%%%%%%%%%%%%%%%%%%%
\subsection{Scenario 5}\label{Scenario5}

Finally, we investigate the scenario where the SU-Tx operates on estimates of the gains $\gp$ and $\gsp$, which would typically arise from the information being fed back via a feedback channel, whose quantization and delay further contributes to the estimation error.  In such a case, we aim to satisfy
\begin{align}\label{eq:PrScen5}
	\textrm{Pr}\left( \Pp  \gp \geq \gT \Ps \gsp + \gT \sss  \biggl| \hgp, \hgsp \right)=1-\alpha,
\end{align}
which must be solved for $\Ps$. The complexity of \eqref{eq:PrScen5} makes it infeasible to derive any analytical capacity results. Instead, \eqref{eq:PrScen5} is derived in Appendix \ref{appScen5} and is shown to be equivalent to
\begin{align}\label{eq:PrScen5idsMain}
	 \sum_{j=0}^\infty \frac{(\lambda_1/2)^j}{j!}e^{-\lambda_1/2}
		&\left(
		1-e^{-\frac{\lambda_2+\beta}{2}} e^{\frac{\lambda_2}{4(\alpha+1)}}
		\sqrt{\frac{8}{\lambda_2(\alpha+1)}} \times \right. \nonumber \\
		& \left. \sum_{r=0}^j \sum_{s=0}^r \left( \frac{\beta}{2}\right)^{r} \frac{\left(\frac{2\alpha}{\beta(\alpha+1)}\right)^{s}}{(r-s)!}
		M_{-s-1/2,0}\left(\frac{\lambda_2}{2(\alpha+1)}\right)
		\right)  = \alpha,
\end{align}
where $M_{\mu,\nu}(\cdot)$ is a Whittaker function. The transmit power, $\Ps$, is then computed using a numerical root finder to solve \eqref{eq:PrScen5idsMain} and the resulting transmit power is used  in capacity simulations. Note that the analysis in Appendix \ref{appScen5} is an essential  part of the simulation scheme since it avoids a simulated search for the $\Ps$ value for every realization of $\hgp$ and $\hgsp$.

%%%%%%%%%%%%%%%%%%%%%%%
\subsection{SU Blocking}\label{SUBlocking}

Using the results derived in Sections \ref{Scenario1} - \ref{Scenario4}, we can derive the SU blocking conditions, that is the probability or condition under which the SU is not allowed to transmit due to the constraint \eqref{eq:SINRconstraint}. 

In the case of \textit{Scenarios} 1 and 2, where $\Ps$ is dependent on the instantaneous value of $\gp$, via \eqref{eq:PsScen1} and \eqref{eq:PsScen2}, respectively, we can compute the probability of SU blocking, by solving for $\textrm{Pr}(\Pt \leq 0)$ or equivalently $\textrm{Pr}(\Ps\leq 0)$. It is easily shown that for \textit{Scenarios} 1 and 2
\begin{align}\label{eq:SUBlock12}
	\textrm{Pr}(\Pt  \leq 0)=1-e^{-\frac{\gT\ssp}{\Op\Pp}}=1-e^{-c_2}.
\end{align}
%
%where we used the definition of $c_2$ in \eqref{eq:c2}.

For \textit{Scenarios} 3 and 4, the SU blocking condition is determined purely from the system parameters, and can be obtained by setting $\Ps \leq 0$ in \eqref{eq:PsScen3} and \eqref{eq:PsScen4}, respectively. Here,  the SU blocking condition is related to $\alpha$ and $c_2$ by
\begin{align}\label{eq:SUBlock34}
	\Pt  = 0 \quad \text{if} \quad \alpha \leq 1-e^{-\frac{\gT\Op}{\ssp \Pp }}=1-e^{-c_2}.
\end{align}
Using \eqref{eq:SUBlock34}, we note that for small values of $\alpha$, that is where we guarantee the PU SINR constraint with high probability, the SU blocking condition is approximated by $\alpha \leq c_2$.

For \textit{Scenario} 5, blocking occurs when \eqref{eq:PrScen5} can not be satisfied, even for $\Ps=0$. Hence, the blocking probability is equivalent to
\begin{align}\label{eq:SUBlock5}
\textrm{Pr}\left\{\textrm{Pr}\left(
		\gp \geq \frac{\gT\ssp}{\Pp} \Biggl|  \hgp
	\right)<1-\alpha\right\}.
\end{align}
Converting the inner probability in \eqref{eq:SUBlock5} to a standard non-central $\chi^2$ probability, we use the variable $X$, in \eqref{eq:XYchi2}, to rewrite \eqref{eq:SUBlock5} as
\begin{align}\label{eq:SUBlock5a}
&\textrm{Pr}\left\{\textrm{Pr}\left(
	X \geq \frac{2\gT\ssp}{\Op(1-\rho^2)\Pp} \Biggl|  \hgp
	\right)<1-\alpha\right\} %\nonumber\\ &
=	\textrm{Pr}\left\{\textrm{Pr}\left(
	X \geq \frac{2c_2}{1-\rho^2} \Biggl|  \hgp
	\right)<1-\alpha\right\} ,
\end{align}
where the dependence of $X$ on $\hgp$ lies in the non-centrality parameter, $\lambda_1=2\rho^2\hgp/(\Op(1-\rho^2))$. In order to evaluate \eqref{eq:SUBlock5a} we solve
\begin{align}\label{eq:SUBlock5b}
\textrm{Pr}\left(
		X \geq \frac{2c_2}{1-\rho^2}  \Biggl|  \hgp
	\right) = 1-\alpha,
\end{align}
by a simple root-finder, to find the threshold value, $\hgp=g^*$, which satisfies \eqref{eq:SUBlock5b}. Then, the blocking probability is simply
\begin{align}\label{eq:SUBlock5c}
\textrm{Pr}\left(
		\hgp < g^*\right)=1-e^{-g^*/\Op}.
\end{align}
%
%%%%%%%%%%%%%%%%%%%%%%%%%%%%%%%%%%%%%%%%%%%%%%%%%%%%%%%%%%
\section{Simulation Results and Discussion}\label{Simulations}
We now present simulation results to validate the analytical expressions derived in Section \ref{SUCapacity}, and to compare capacity values achievable under each scenario.  In all simulations, we have set $\Pp /\ssp  = \Pm /\sss  = 0$ dB and $\Op /\ssp =\Os /\sss = 5$ dB, where we assume $\ssp =\sss$. In \textit{Scenarios} 2-5 we set $\alpha=0.1$, and $\rho=0.9$  is used in \textit{Scenario} 5, unless otherwise indicated in the figures.

\begin{figure}[e]
	\centering \incgraphicswidth{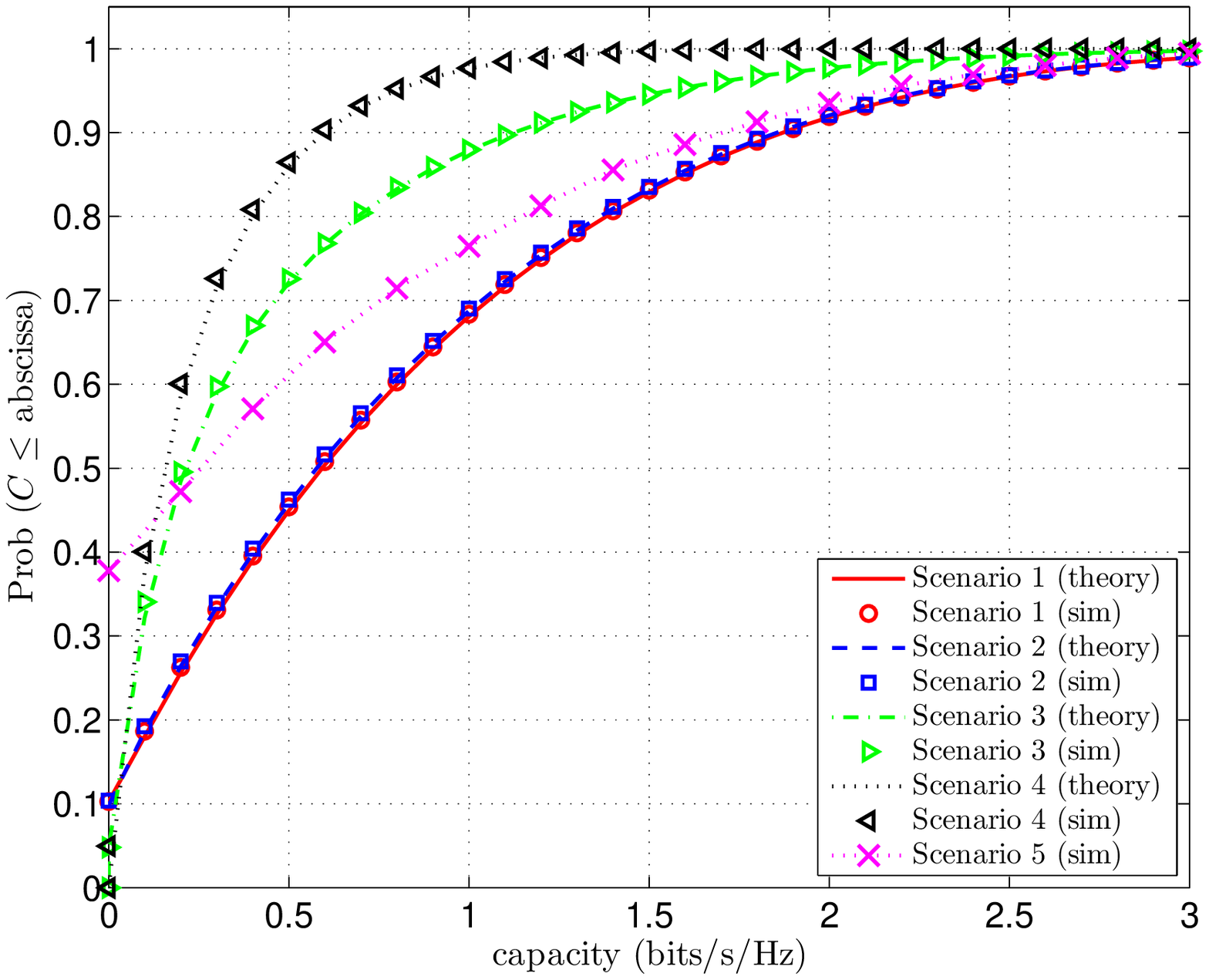}
	\raisecaption\caption{SU capacity cdf for \textit{Scenarios} 1-4 ($c_1=c_2=0.1$).}
	\label{figCapCDF1234Inta}
\end{figure}

\begin{figure}[e]
	\centering \incgraphicswidth{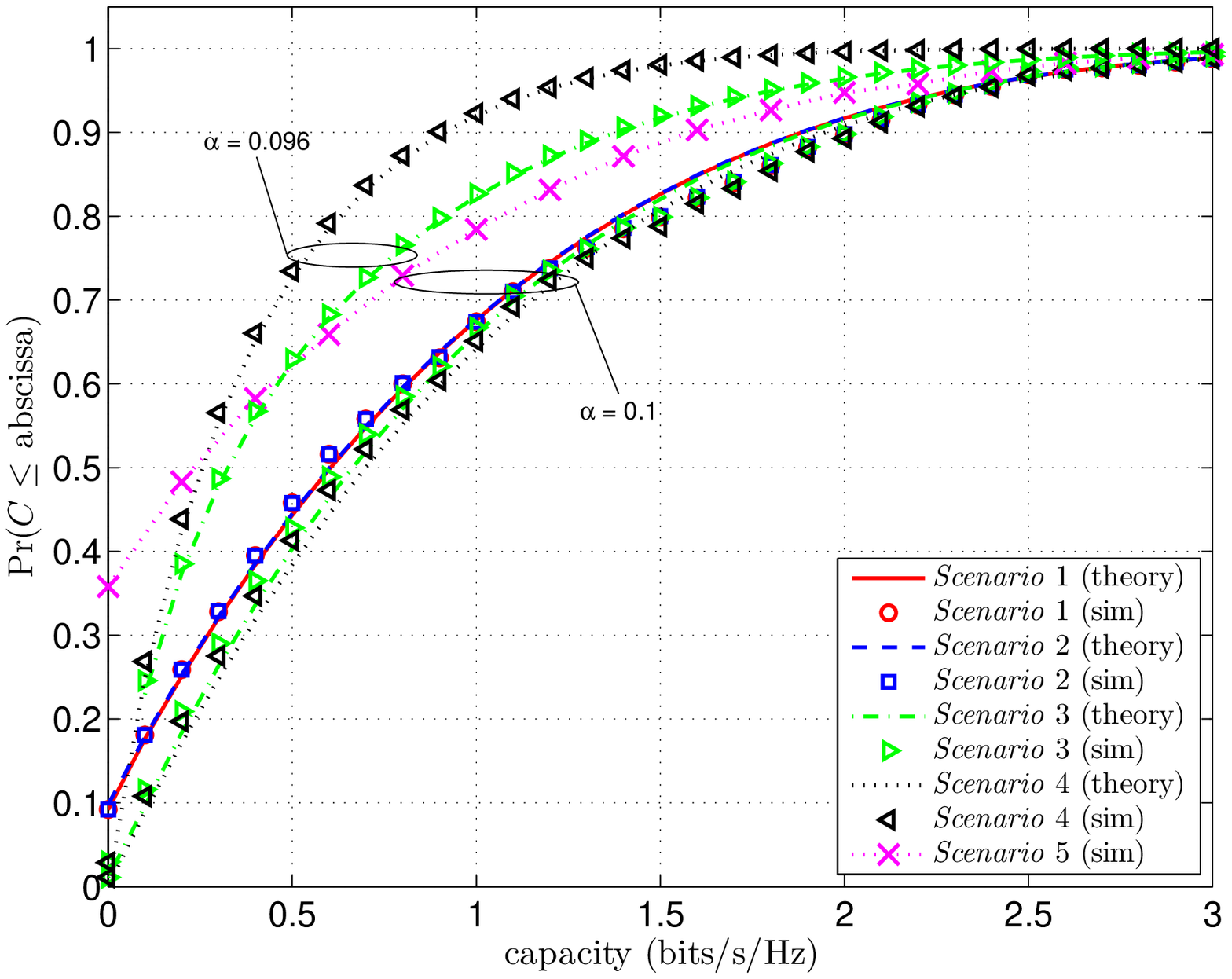}
	\raisecaption\caption{SU capacity cdf for \textit{Scenarios} 1-4 ($c_1=0.01, c_2=0.1$).}
	\label{figCapCDF1234Intaa}
\end{figure}

\begin{figure}[e]
	\centering \incgraphicswidth{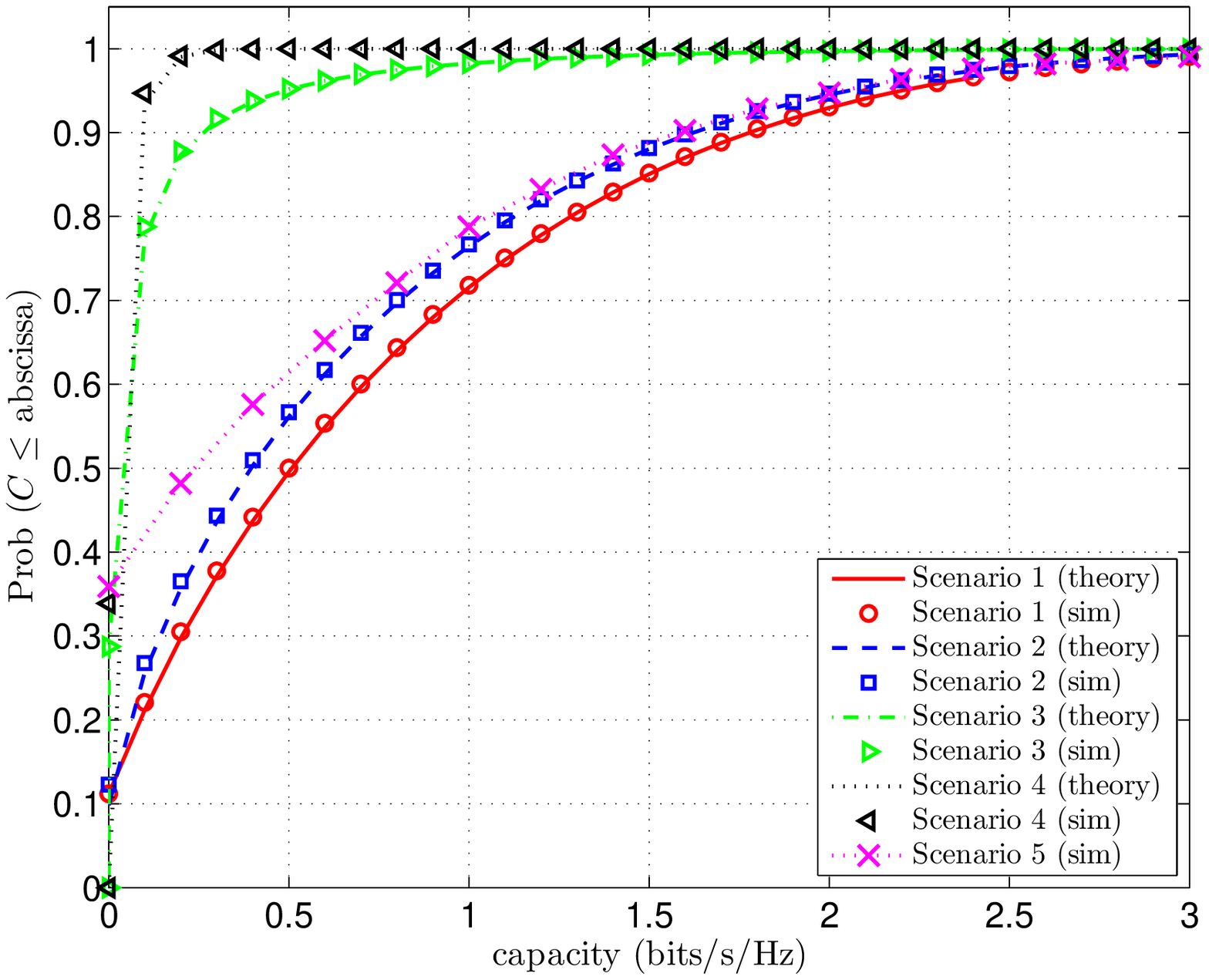}
	\raisecaption\caption{SU capacity cdf for \textit{Scenarios} 1-4 ($c_1=0.9, c_2=0.1$).}
	\label{figCapCDF1234Intb}
\end{figure}

\begin{figure}[e]
	\centering \incgraphicswidth{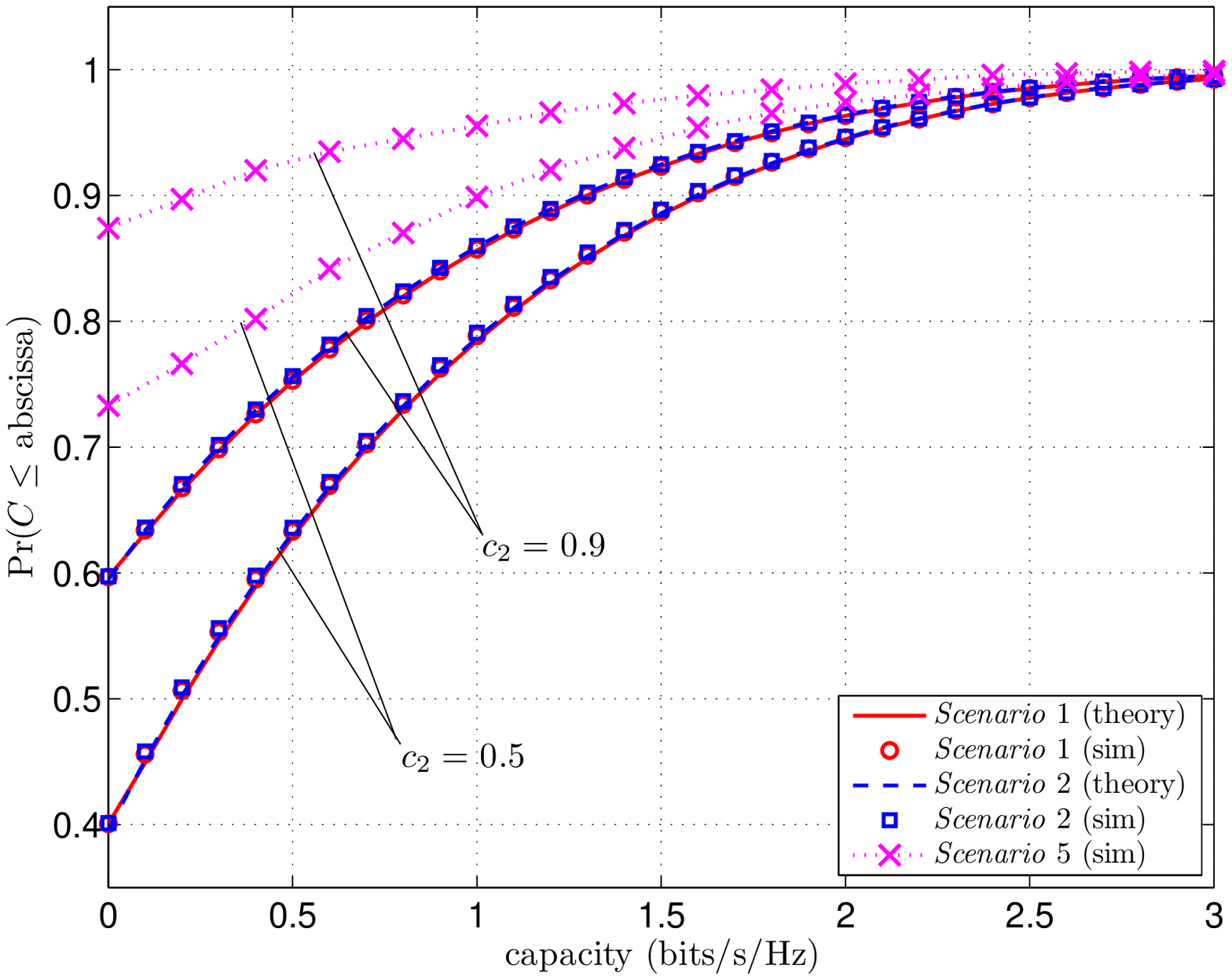}
	\raisecaption\caption{SU capacity cdf for \textit{Scenarios} 1 and 2 ($c_1=0.01$; $c_2=0.5$, $c_2=0.9$)}
	\label{figCapCDF1212Int}
\end{figure}

Figures \ref{figCapCDF1234Inta}, \ref{figCapCDF1234Intaa}, \ref{figCapCDF1234Intb} and \ref{figCapCDF1212Int} show the SU capacity cdfs for various scenarios and a range of $c_1$, $c_2$ values.  Figures \ref{figCapCDF1234Intaa} and \ref{figCapCDF1212Int}, with $c_1=0.01$, represent very favourable SU operating conditions due to either relative distances or the levels of shadowing being experienced by the PU and SU receivers.  Figures \ref{figCapCDF1234Inta} and \ref{figCapCDF1234Intb} ($c_1=0.1$ and $c_1=0.9$) represent increasingly difficult conditions for the SU. The analytical expressions were obtained using \eqref{eq:FgIScen1}, \eqref{eq:FgIScen2} or \eqref{eq:FgIScen3} substituted into \eqref{eq:FCgenI}, and \eqref{eq:FCintScen4} for \textit{Scenarios} 1-3 and \textit{Scenario} 4, respectively. Results for \textit{Scenario} 5 were obtained by solving \eqref{eq:PrScen5idsMain} for SU power $\Ps$, restricted by \eqref{eq:Pt}, and substituting $\Pt$ into \eqref{eq:instantC} via \eqref{eq:gammaI}.
\begin{figure}[e]
	\centering \incgraphicswidth{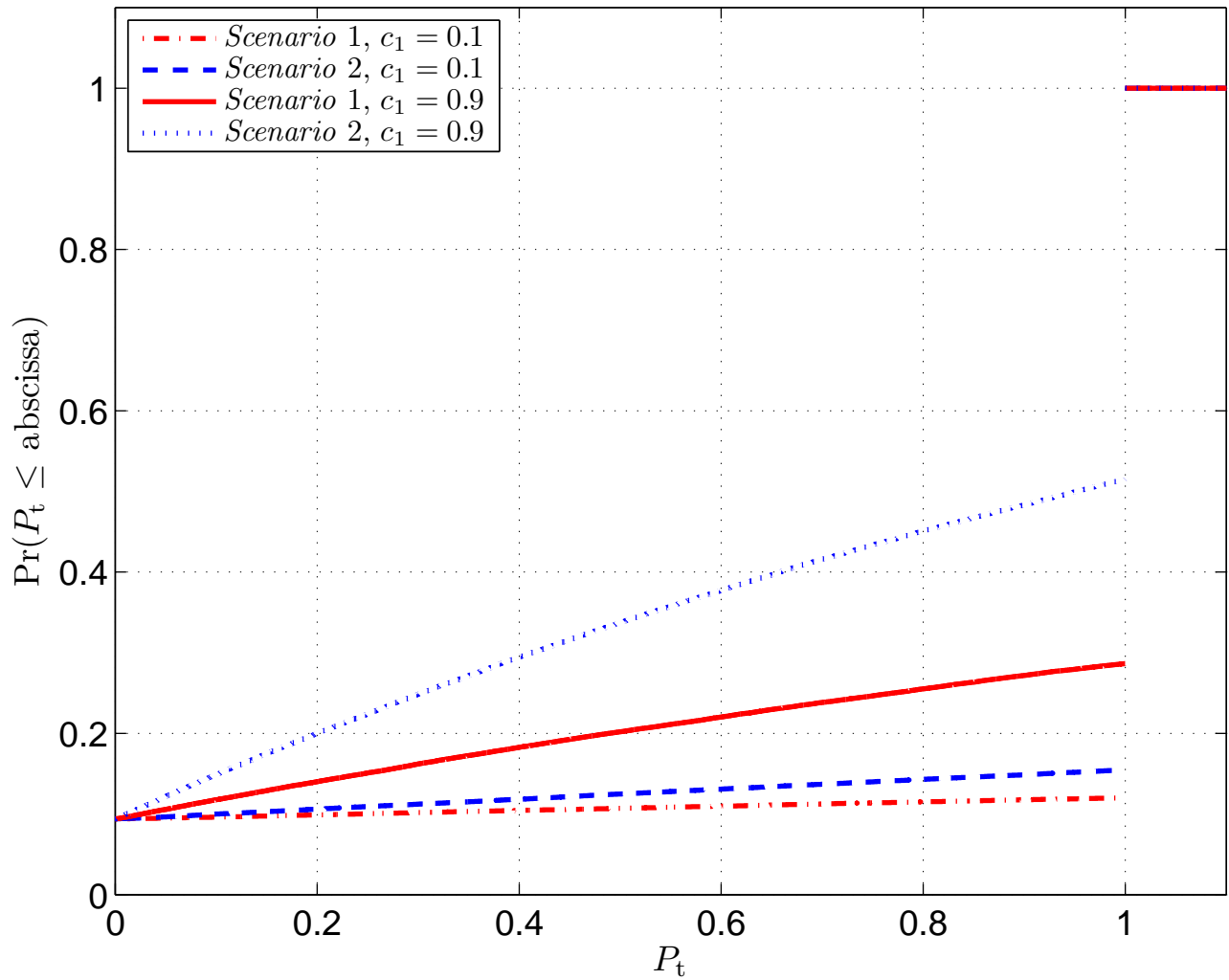}
	\raisecaption\caption{SU transmit power cdf for \textit{Scenarios} 1 and 2 ($c_1=0.1$ and $c_1=0.9$).}
	\label{figPsCDF12}
\end{figure}

From these results, we observe that \textit{Scenarios} 1 and 2 result in similar performance, even in the case of $c_1=0.9$ (Fig. \ref{figCapCDF1234Intb}), that is where the SU interference is very prominent. %, thus making it difficult to satisfy the PU SINR constraint in \eqref{eq:SINRconstraint}. 
Furthermore, lack of knowledge of the PU-PU link (that is, knowing only the mean $\Op $) greatly reduces the achievable capacity of the SU. This is shown in Figs \ref{figCapCDF1234Inta}, \ref{figCapCDF1234Intaa} and \ref{figCapCDF1234Intb} where \textit{Scenarios} 3 and 4 suffer a considerable  loss in comparison to \textit{Scenarios} 1 and 2. Hence, knowledge of $\gp$ is more important than knowledge of $\gsp$. %From the point of view of any SU-PU collaboration, this is a positive result, since $\gp$ is likely to be known at the PU-Rx. Hence, the PU already has the most useful information and does not need to estimate $\gsp$, which might make the price of co-existence too high from the PU perspective.

The dependence on $c_1$ can be observed by comparing Figs. \ref{figCapCDF1234Inta}, \ref{figCapCDF1234Intaa} and \ref{figCapCDF1234Intb}. Under very favourable conditions, $c_1=0.01$, \textit{Scenarios} 3 and 4 slightly outperform \textit{Scenarios} 1 and 2. This seemingly counterintuitive result is due to the flexibility afforded by the probabilistic SINR constraint. This is confirmed by the additional cdfs for \textit{Scenarios} 3 and 4 in Fig. \ref{figCapCDF1234Intb}, with $\alpha =0.096$, where the protection of PU SINR with higher degree of certainty causes degradation of performance for \textit{Scenarios} 3 and 4 below that for \textit{Scenarios} 1 and 2. The high sensitivity to the parameter $\alpha$ is due to the fact that the SU is operating near the blocking condition given by \eqref{eq:SUBlock34}, which for $c_2=0.1$ requires $\alpha > 0.0952$ in order to allow SU transmission.

From Fig. \ref{figCapCDF1234Intb}, we observe that placing the SU in a demanding environment, $c_1=0.9$, results in very poor performance under \textit{Scenarios} 3 and 4. Furthermore, the performance of \textit{Scenario} 2 is noticeably degraded from that of \textit{Scenario} 1. Further insight into this is provided by Fig. \ref{figPsCDF12}, which shows the cdf of the SU transmit power, $\Pt$, for $c_1=0.1$ and $c_1=0.9$. We observe that in the latter case, the SU-Tx under \textit{Scenario} 1 operates at maximum power, $\Pt=1$, with a likelihood of 70 \%, compared to approximately 50 \% for \textit{Scenario} 2. This difference is much less pronounced for the less challenging case of $c_1=0.1$.  Finally, based on Figs. \ref{figCapCDF1234Inta}, \ref{figCapCDF1234Intaa} and \ref{figCapCDF1234Intb}, we observe that the performance under \textit{Scenario} 5 is not highly dependent on the value of $c_1$.

Comparing the curves for \textit{Scenarios} 3 and 4 with those for \textit{Scenario} 5 in Fig. \ref{figCapCDF1234Intb} we note that for the most part % for moderate capacity regions 
imperfect knowledge of the channel gains is more beneficial to the knowledge of their mean. Only in low capacity regime we observe that \textit{Scenarios} 3 and 4 outperform \textit{Scenario} 5, which has a relatively high blocking probability for the parameters considered. It should be noted, however, that blocking in \textit{Scenarios} 3 and 4 is dictated by the parameter $c_2$ and thus, unless \eqref{eq:SUBlock34} is satisfied, the capacity cdfs for these scenarios originate at zero. Consequently, at higher capacity values there exists a crossover point with \textit{Scenario} 5.

Figures \ref{figCapCDF1234Inta}, \ref{figCapCDF1234Intaa} and \ref{figCapCDF1234Intb} compare the scenarios using $c_2=0.1$, which is very generous to the SU. From \eqref{eq:SUBlock34}, we see that SU transmission in \textit{Scenarios} 3 and 4 occurs only for large values of $\alpha$ or for small values of $c_2$. That is, without the knowledge of $\gp$, the SU can only operate if the PU is willing to accept large amounts of interference. Figure \ref{figCapCDF1212Int} presents the capacity results for \textit{Scenarios} 1 and 2 with the more realistic values of $c_2=0.5$ and $c_2=0.9$, where \eqref{eq:SUBlock34}  prevents SU transmission under \textit{Scenarios} 3 and 4. While SU transmission is possible under \textit{Scenario} 5, we observe a high blocking probability of 0.73 and 0.88 for $c_2=0.5$ and $c_2=0.9$, respectively.

\begin{figure}[e]
	\centering \incgraphicswidth{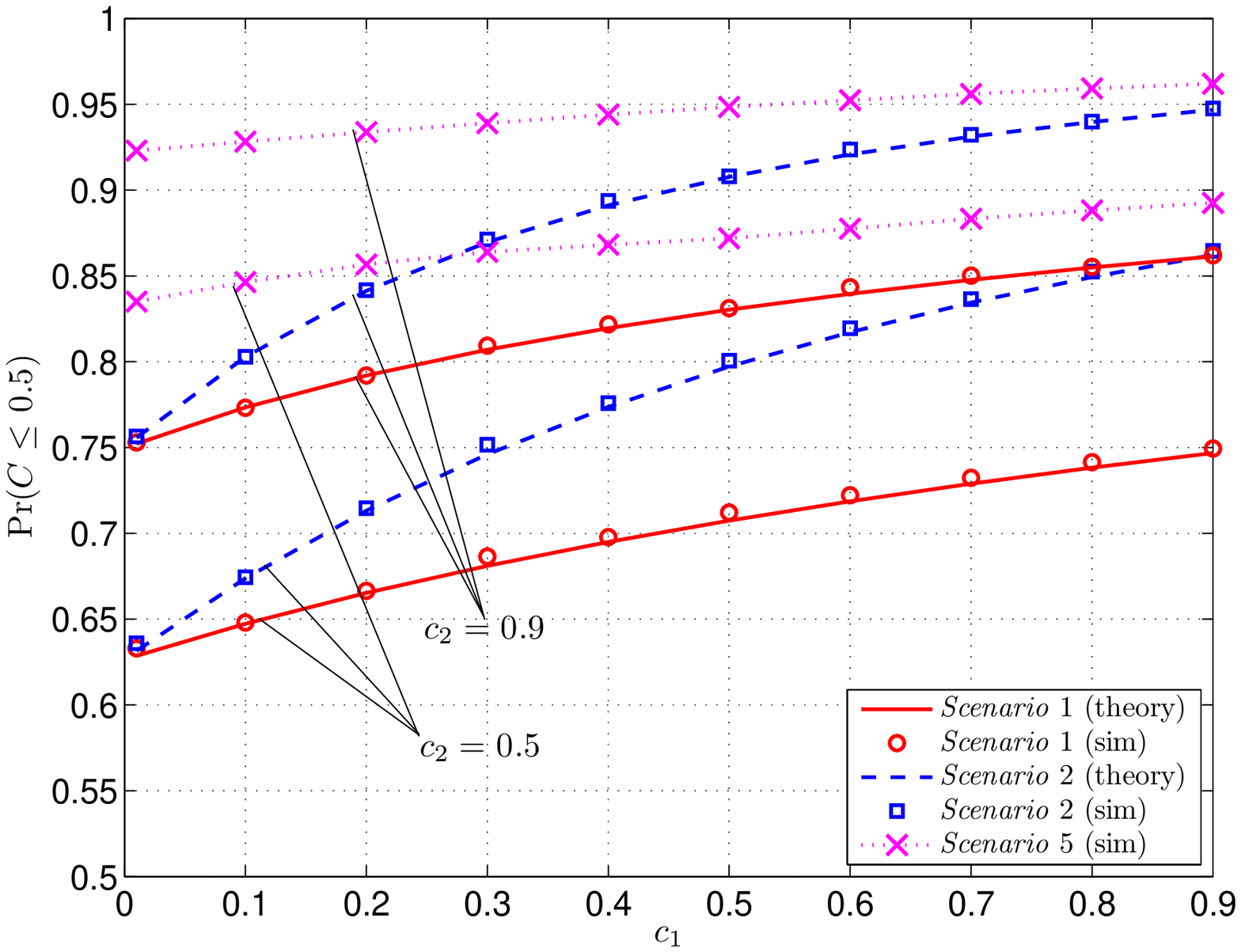}
	\raisecaption\caption{SU capacity outage values for \textit{Scenarios} 1 and 2 vs $c_1$ ($c_2=0.5$, $c_2=0.9$).}
	\label{figCapCDF1212Intc1}
\end{figure}

Figure \ref{figCapCDF1212Intc1} shows the probability $\textrm{Pr}(C \leq 0.5)$ as a function of $c_1$. As expected, for a constant $c_2$, the performance under \textit{Scenario} 2 diverges from the baseline \textit{Scenario} 1 with increasing $c_1$, that is as the amount of interference to the PU increases.

\begin{figure}[e]
	\centering \incgraphicswidth{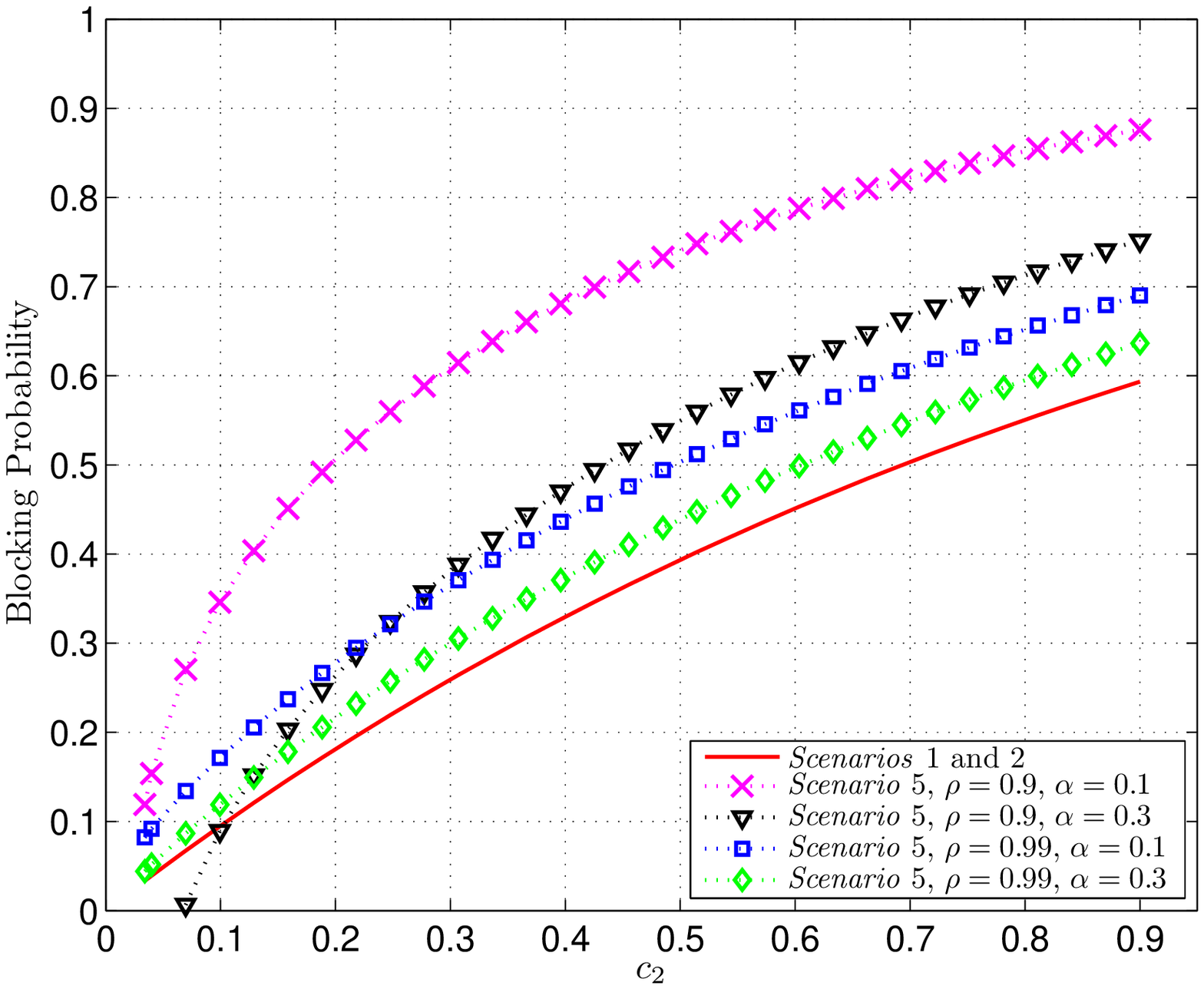}
	\raisecaption\caption{Blocking probability for \textit{Scenarios} 1, 2 and 5 for vs $c_2$.}
	\label{figBlockProb}
\end{figure}

Finally, Fig.	\ref{figBlockProb} shows the blocking probability for \textit{Scenarios} 1, 2 and 5. We recall that the SU ability to transmit in \textit{Scenarios} 3 and 4 is deterministic and governed by the blocking condition of \eqref{eq:SUBlock34}. The results for \textit{Scenario} 5 were obtained numerically via \eqref{eq:SUBlock5b}.
We observe that, as expected, as the channel knowledge error decreases ($\rho \rightarrow 1$) the blocking probability approaches that of \textit{Scenarios} 1 and 2. Specifically, referring back to Fig. \ref{figCapCDF1212Int}, we note from Fig. \ref{figBlockProb} that improving the channel estimate to $\rho=0.99$ will reduce the blocking probability at $c_2=0.5$ and $c_2=0.9$ to 0.5 and 0.7, respectively, thus bringing the performance of \textit{Scenario} 5 closer to that of \textit{Scenario} 1. Similarly, relaxing the probabilistic SINR constraint by increasing $\alpha$ to 0.3 results in a significant reduction in blocking probability, as fully expected.
%
%%%%%%%%%%%%%%%%%%%%%%%%%%%%%%%%%%%%%%%%%%%%%%%%%
\section{Conclusions}
%
%In this paper, w
We have examined the effects of limited channel knowledge on the SU capacity. Considering five scenarios, we derived (in four cases) analytical expressions for the SU capacity cdf under an PU-Rx SINR constraint. We determined the SU blocking probability and blocking conditions as a function of the allowable interference at the PU-Rx. The results demonstrate the importance of the PU-PU CSI, which was shown to be much greater than that of the SU-Tx to PU-Rx link. %This can be seen as a positive result, as the PU-Rx is not required to obtain a precise estimate of this additional link to enable SU transmission. 
Furthermore, we have shown that in challenging situations or in the presence of CSI error there can be extremely large blocking probabilities for the SUs.
%
%%%%%%%%%%%%%%%%%%%%%%%%%%%%%%%%%%%%%%%%%%%%%%%%%
\appendices
\section{Derivation of $F_{\gamI}(\ty)$ and $f_{\gamI}(\ty)$ for \textit{Scenario} 3}
\label{appScen3}
We now derive the SU SINR cdf $F_{\gamI}(\ty)$ and pdf $f_{\gamI}(\ty)$, based on the SU transmit power $\Ps$ under the SINR constraint of \textit{Scenario 3}, given by \eqref{eq:PsScen3}. Defining for notational convenience
\begin{align}\label{eq:QScen3}
	Q=-\left(\frac{\ln(1-\alpha)\Pp \Op }{\gT }+\ssp \right),
\end{align}
the transmit power for \textit{Scenario} 3 is given by
\begin{align}\label{eq:PtScen3}
	\Pt =\min\left(\Pm ,\frac{Q}{\gsp}\right).
\end{align}
Solving for the cdf of $\gamma$, we write
\begin{align}\label{eq:PrgScen3aa}
		\textrm{Pr} \left(\gamma>x\right)  & = 	\textrm{Pr} \left(\Pm \gs >x, \Ps \gs>x \right) \nonumber \\
		&=\mathbb{E}\left[ \textrm{Pr}\left( \Pm \gs > x, \Ps \gs>x \Bigl| \gsp  \right)   \right],
\end{align}
where the conditional probability in \eqref{eq:PrgScen3aa} is given by
\begin{align}\label{eq:PrgScen3a}
		 \textrm{Pr}\left( \Pm \gs > x, \Ps \gs>x \Bigl| \gsp  \right)
		= \begin{cases}	\textrm{Pr}\left(\gs  > \frac{x}{\Pm} \right) & \Pm  \leq \frac{Q}{\gsp} \\
										\textrm{Pr}\left(\gs  > \frac{x}{\Ps} \right) & \Pm  > \frac{Q}{\gsp}.
			\end{cases}
\end{align}
Hence, we have
\begin{align}\label{eq:PrgScen3b}
		\textrm{Pr} \left(\gamma>x\right)
		& = 	\int_{\psi_0}^{Q/\Pm }\textrm{Pr}\left(\gs  > \frac{x}{\Pm } \right) f_{\gsp}(y) \ dy % \\ \nonumber &
		 + \int_{Q/\Pm }^{\infty}\textrm{Pr}\left(\gs  > \frac{xy}{Q} \right) f_{\gsp}(y) \ dy.
\end{align}
Upon simplifying \eqref{eq:PrgScen3b}, we obtain
\begin{align}\label{eq:FgScen3}
		F_{\gamma}(x) & =1-e^{-\frac{x}{\Osp \Pm }} \left(1-e^{-\frac{Q}{\Os \Pm }}\right) %\\ \nonumber &
									-\frac{1}{1+\frac{\Os }{\Osp  Qx} } e^{-\frac{Q}{\Osp  \Pm }-\frac{x}{\Os  \Pm }}.
\end{align}
Substituting \eqref{eq:FgScen3} into \eqref{eq:FgIgen} and after some manipulation one obtains
\begin{align}\label{eq:FgIScen3}
		F_{\gamma_I}(\tilde{y}) & = 1-\frac{\left( 1-e^{-Q/\Pm \Osp } \right)}{1+\frac{\ty \Pp \Ops }{\Pm \Os }}e^{-\ty \sss /\Pm \Os }+I(\ty),
\end{align}
where $I(\ty)$ is given by
\begin{align}\label{eq:FgIScen3Int}
		I(\ty)&=-\frac{Q\Os}{\Ops\Osp}
			\exp\left\{ -\frac{Q\Os+\sss\ty}{\Pm \Os\Osp} \right\}	
		 \int_{0}^{\infty}
		{\frac{ \exp\left\{  -\left( \frac{\Pp \ty}{\Pm \Os}+\frac{1}{\Ops}  \right)v \right\} }
		{\sss \ty +\frac{Q\Os}{\Osp} + \Pp \ty v} \ dv}.
\end{align}
The integral in \eqref{eq:FgIScen3Int} can be solved using \cite[Eq. (3.352.4)]{Gradshteyn2007}, resulting in
\begin{align}\label{eq:FgIScen3Int2}
		I(\ty)&=-\frac{Q\Os}{\Ops\Osp \Pp \ty}
		\exp\left\{ -\frac{Q\Os+\Osp\sss\ty}{\Pm \Os\Osp} \right\}				
		\exp\left\{ \frac{(\Pp \Ops\ty+\Pm \Os)(\sss\Osp\ty+Q\Os)}{\Pm \Pp \Os\Ops\Osp\ty} \right\}	 \\ \nonumber
				&\times \textrm{E}_1\left(
				\frac{(\Pp \Ops\ty+\Pm \Os)(\sss\Osp\ty+Q\Os)}{\Pm \Pp \Os\Ops\Osp\ty} \right),
\end{align}
where $\textrm{E}_1(\cdot)$ denotes the exponential integral. Equation \eqref{eq:FgIScen3} can be expressed in terms of simple functions of $\ty$ by
%%
%\begin{align}\label{eq:FgIScen3comp}
		%F_{\gamma_I}(\tilde{y}) & = 1-s(\ty) -h(\ty) \textrm{E}_1(r(\ty)),
%\end{align}
%%
%%
%as given in 
\eqref{eq:FgIScen3compMain}, %
where
\begin{align}\label{eq:shr}
s(\ty)&=\frac{K_1 e^{-b \ty}}{1+a\ty}, % \nonumber\\
\quad \quad
h(\ty)=\frac{K_2e^{-b\ty+r(\ty)}}{\ty}, \\
r(\ty)&=\frac{(\Pp \Ops\ty+\Pm \Os)(\sss\Osp\ty+Q\Os)}{\Pm \Pp \Os\Ops\Osp\ty}, \nonumber
\end{align}
with constants, $K_1=1-e^{Q/\Pm \Osp}$, $K_2=\frac{Q\Os e^{Q/\Pm \Osp}}{\Pp \Ops\Osp}$, $a=\frac{Pp\Ops}{\Pm \Os}$ and $b=\sss/\Pm \Os$. Taking the derivative of \eqref{eq:FgIScen3compMain} gives
%%
%\begin{align}\label{eq:fgIScen3comp}
		%f_{\gamma_I}(\tilde{y}) & = -s'(\ty)-h'(\ty) \textrm{E}_1(r(\ty))+h(\ty)r'(\ty)\frac{e^{-r(\ty)}}{r(\ty)},
%\end{align}
%%
%%
%as given in 
\eqref{eq:fgIScen3compMain}, %
where the derivatives of $s(\ty)$, $h(\ty)$ and $r(\ty)$
%the functions in \eqref{eq:shr}  
are given by
\begin{align}\label{eq:dershr}
s'(\ty)&= -\left( \frac{K_1 a +bK_1(1+a\ty) }{(1+a\ty)^2} \right) e^{-b\ty},  \nonumber \\
%h'(\ty)&= -\frac{K_2 e^{-b\ty+r(\ty)}}{\ty^2}+\frac{K_2}{\ty} \left( (-b\ty+r(\ty)) (-b+r'(\ty))e^{-b\ty+r(\ty)}           \right) = below ,  \\
h'(\ty)&= \left( \frac{\ty(r(\ty)-b\ty) (r'(\ty)-b)K_2-K_2} {\ty^2}  \right)  e^{-b\ty+r(\ty)  } ,     \\
r'(\ty)&= \frac{\Pp \sss\Ops\Osp \ty^2 - Q\Pm \Os^2}{(\Pm \Pp \Os\Ops\Osp\ty)^2}. \nonumber
\end{align}
%
%%%%%%%%%%%%%%%%%%%%%%%%%%%%%%%%%%%%%%%%%%%%%%%%
\section{Computation of $\Ps$ for \textit{Scenario} 5}
\label{appScen5}
We now derive the expression used to numerically compute the SU power, $\Ps$, for \textit{Scenario} 5, based on the SINR constraint given by \eqref{eq:PrScen5}. %
Consider the standard estimation error model \cite{Ahn2009}
\begin{align}\label{eq:hest}
	h_p&=\rho \hhp +\sqrt{1-\rho^2} \tep, %\\ \notag
	\quad \quad
	h_{sp}=\rho \hhsp +\sqrt{1-\rho^2} \tesp,
\end{align}
where $\tep$ and $\tesp$ are unit variance complex, zero mean Gaussian random variables and $\rho$ controls the accuracy of the estimates. We assume equal $\rho$ for $h_p$ and $h_{sp}$. Defining
\begin{align}\label{eq:esterr}
	e_p&= \tep \sqrt{1-\rho^2}, % \\ \notag
	\quad \quad
	e_{sp}= \tesp \sqrt{1-\rho^2}, %\notag
\end{align}
gives
\begin{align}\label{eq:PrScen5b}
	\textrm{Pr}\left(  \bigl| \rho \hhp+e_p  \bigl|^2 \geq \frac{  \gT \Ps   \bigl| \rho \hhsp+e_{sp}  \bigl|^2  +\sss\gT }{\Pp }                               \biggl| \hhp, \hhsp  \right)     =1-\alpha.
\end{align}
We now define two non-central $\chi^2$ random variables \cite[p.451-452]{Kotz1994}
\begin{align}\label{eq:XYchi2}
	X &= \frac{2\bigl| \rho \hhp+e_p  \bigl|^2}{\Op(1-\rho^2)}
	\sim  \chi_2^{'2}\left( \frac{2\rho^2\bigl| \hhp \bigl|^2}{\Op(1-\rho^2)}  \right)
	=\chi_2^{'2}\left( \lambda_1 \right), \\ \nonumber
	Y &= \frac{2\bigl| \rho \hhsp+e_{sp}  \bigl|^2}{\Osp(1-\rho^2)}
	\sim  \chi_2^{'2}\left( \frac{2\rho^2\bigl| \hhsp\bigl|^2}{\Osp(1-\rho^2)}  \right)=\chi_2^{'2}\left( \lambda_2 \right),
\end{align}
and with this notation, \eqref{eq:PrScen5b} becomes
\begin{align}\label{eq:PrScen5c}
	\textrm{Pr}\left( X \geq \frac{\gT \Ps \Osp}{\Op \Pp } Y + \frac{2\sss \gT}{\Op(1-\rho^2)\Pp }
										\bigl| \hhp, \hhsp \right)     =1-\alpha
\end{align}
or
\begin{align}\label{eq:PrScen5d}
	\textrm{Pr}\left( X \geq \alpha Y + \beta
										\bigl| \hhp, \hhsp \right)     =1-\alpha.
\end{align}
Dropping the conditioning notation for ease of exposition and conditioning on $e_{sp}$ (equivalent to conditioning on $Y$) results in
\begin{align}\label{eq:PrScen5e}
	\mathbb{E}\left[ \textrm{Pr}\left( X \geq \alpha Y + \beta
										\bigl| Y\right)   \right]  & =1-\alpha \\ \nonumber
\Rightarrow	\mathbb{E}\left[ \textrm{Pr}\left( X \leq \alpha Y + \beta
										\bigl| Y\right)   \right]  & =\alpha.
\end{align}
Using the cdf of a non-central $\chi^2$ in \eqref{eq:PrScen5e} gives \cite{Kotz1994}
\begin{align}\label{eq:PrScen5f}
	\mathbb{E}\left[ \sum_{j=0}^\infty \frac{(\lambda_1/2)^j}{j!}e^{-\lambda_1/2}
		\left(  1-e^{-(\alpha Y +\beta)/2} \sum_{r=0}^j \left( \frac{\alpha Y + \beta}{2}  \right)^r\right)
	  \right] = \alpha.
\end{align}
Expanding the binomial series in \eqref{eq:PrScen5f} gives
\begin{align}\label{eq:PrScen5g}
	\mathbb{E}\left[ \sum_{j=0}^\infty \frac{(\lambda_1/2)^j}{j!}e^{-\lambda_1/2}
		\left(  1-e^{\beta/2} \sum_{r=0}^j \sum_{s=0}^r {r \choose s} \left( \frac{\beta}{2}\right)^{r-s} \left(\frac{\alpha}{2}\right)^{s}  Y^s e^{-\alpha Y/2} \right) \right] = \alpha.
\end{align}
Taking expectation over Y using the pdf of a non-central $\chi^2$ \cite{Kotz1994} gives
\begin{align}\label{eq:PrScen5h}
	 \sum_{j=0}^\infty \frac{(\lambda_1/2)^j}{j!}e^{-\lambda_1/2}
		\left(  1-e^{\beta/2} \sum_{r=0}^j \sum_{s=0}^r {r \choose s} \left( \frac{\beta}{2}\right)^{r-s} \left(\frac{\alpha}{2}\right)^{s} J(s) \right)  = \alpha,
\end{align}
where
\begin{align}\label{eq:PrScen5Js}
	 J(s)=\int_{0}^{\infty}y^s e^{-\alpha y /2} I_0(\sqrt{y\lambda_2}) e^{(y+\lambda_2)/2} \ dy,
\end{align}
where $I_0(\cdot)$ is the zeroth order modified Bessel function of the first kind. Using the result in \cite[Eq. (6.643)]{Gradshteyn2007} gives after simplification the final result in \eqref{eq:PrScen5idsMain}.
%%
%\begin{align}\label{eq:PrScen5ids}
	 %\sum_{j=0}^\infty \frac{(\lambda_1/2)^j}{j!}e^{-\lambda_1/2}
		%&\left(
		%1-e^{-\frac{\lambda_2+\beta}{2}} e^{\frac{\lambda_2}{4(\alpha+1)}}
		%\sqrt{\frac{8}{\lambda_2(\alpha+1)}} \times \right. \nonumber \\
		%& \left. \sum_{r=0}^j \sum_{s=0}^r \left( \frac{\beta}{2}\right)^{r} \frac{\left(\frac{2\alpha}{\beta(\alpha+1)}\right)^{s}}{(r-s)!}
		%M_{-s-1/2,0}\left(\frac{\lambda_2}{2(\alpha+1)}\right)
		%\right)  = \alpha
%\end{align}
%%
%where $M_{\mu,\nu}(\cdot)$ is a Whittaker function.

\bibliographystyle{IEEEtran}
\bibliography{TVT-CR}

\begin{thebibliography}{10}
\providecommand{\url}[1]{#1}
\csname url@samestyle\endcsname
\providecommand{\newblock}{\relax}
\providecommand{\bibinfo}[2]{#2}
\providecommand{\BIBentrySTDinterwordspacing}{\spaceskip=0pt\relax}
\providecommand{\BIBentryALTinterwordstretchfactor}{4}
\providecommand{\BIBentryALTinterwordspacing}{\spaceskip=\fontdimen2\font plus
\BIBentryALTinterwordstretchfactor\fontdimen3\font minus
  \fontdimen4\font\relax}
\providecommand{\BIBforeignlanguage}[2]{{%
\expandafter\ifx\csname l@#1\endcsname\relax
\typeout{** WARNING: IEEEtran.bst: No hyphenation pattern has been}%
\typeout{** loaded for the language `#1'. Using the pattern for}%
\typeout{** the default language instead.}%
\else
\language=\csname l@#1\endcsname
\fi
#2}}
\providecommand{\BIBdecl}{\relax}
\BIBdecl

\bibitem{Mitola2000}
I.~J.~Mitola, ``Cognitive radio: An integrated agent architecture for software
  defined radio,'' Ph.D. dissertation, KTH Royal Institute of Technology,
  Sweden, May 2000.

\bibitem{Haykin2005}
S.~Haykin, ``Cognitive radio: Brain-empowered wireless communications,''
  \emph{IEEE J. Select. Areas Commun.}, vol.~23, pp. 201--220, February 2005.

\bibitem{Weiss2004}
T.~A. Weiss and F.~K. Jondral, ``Spectrum pooling: An innovative strategy for
  the enhancement of spectrum efficiency,'' \emph{IEEE Commun. Mag.}, vol.~42,
  pp. 8--14, March 2004.

\bibitem{Wang2011}
B.~Wang and J.~K.~R. Liu, ``Advances in cognitive radio networks: A survey,''
  \emph{IEEE J. Select. Topics Signal Process.}, vol.~5, pp. 5--23, February
  2011.

\bibitem{Shin2010}
K.~G. Shin, H.~Kim, A.~W. Min, and A.~Kumar, ``Cognitive radios for dynamic
  spectrum access: From concept to reality,'' \emph{IEEE Wireless Commun.
  Mag.}, vol.~17, pp. 64--74, December 2010.

\bibitem{Ghasemi2007}
A.~Ghasemi and E.~S. Sousa, ``Fundamental limits of spectrum-sharing in fading
  environments,'' \emph{IEEE Trans. Wireless Commun.}, vol.~6, pp. 649--658,
  February 2007.

\bibitem{Jafar2007}
S.~A. Jafar and S.~Srinivasa, ``Capacity limits of cognitive radio with
  distributed and dynamic spectral activity,'' \emph{IEEE J. Select. Areas
  Commun.}, vol.~25, pp. 529--537, April 2007.

\bibitem{Musavian2009}
L.~Musavian and S.~Aissa, ``Capacity and power allocation for spectrum sharing
  communications in fading channels,'' \emph{IEEE Trans. Wireless Commun.},
  vol.~8, pp. 148--156, January 2009.

\bibitem{Suraweera2008}
H.~A. Suraweera, J.~Gao, P.~J. Smith, M.~Shafi, and M.~Faulkner, ``Channel
  capacity limits of cognitive radio in asymmetric fading environments,'' in
  \emph{Proc. IEEE International Conference on Communications, 2008. ICC '08.},
  May 2008, pp. 4048--4053.

\bibitem{Zhang2008}
R.~Zhang, ``Optimal power control over fading cognitive radio channels by
  exploiting primary user {CSI},'' in \emph{Proc. IEEE Global
  Telecommunications Conference, 2008. GLOBECOM '08.}, November 2008, pp. 1--5.

\bibitem{Kang2009}
X.~Kang, Y.-C. Liang, A.~Nallanathan, H.~K. Garg, and R.~Zhang, ``Optimal power
  allocation for fading channels in cognitive radio networks: Ergodic capacity
  and outage capacity,'' \emph{IEEE Trans. Wireless Commun.}, vol.~8, pp.
  940--950, February 2009.

\bibitem{Wang2009}
C.-X. Wang, X.~Hong, H.-H. Chen, and J.~Thompson, ``On capacity of cognitive
  radio networks with average interference power constraints,'' \emph{IEEE
  Trans. Wireless Commun.}, vol.~8, pp. 1620--1625, April 2009.

\bibitem{Aissa2009}
L.~Musavian and S.~Aissa, ``Fundamental capacity limits of cognitive radio in
  fading environments with imperfect channel information,'' \emph{IEEE Trans.
  Commun.}, vol.~57, pp. 3472--3480, November 2009.

\bibitem{ShafiTVT2010}
H.~A. Suraweera, P.~J. Smith, and M.~Shafi, ``Capacity limits and performance
  analysis of cognitive radio with imperfect channel knowledge,'' \emph{IEEE
  Trans. Veh. Technol.}, vol.~59, pp. 1811--1822, May 2010.

\bibitem{Alouini2011}
Z.~Rezki and M.-S. Alouini, ``On the capacity of cognitive radio under limited
  channel state information over fading channels,'' in \emph{Proc. 2011 IEEE
  International Conference on Communications (ICC)}, June 2011, pp. 1--5.

\bibitem{AlouiniPIRMC2011}
L.~Sboui, Z.~Rezki, and M.~Alouini, ``Capacity of cognitive radio under
  imperfect secondary and cross link channel state information,'' in
  \emph{Proc. 2011 IEEE 22nd International Symposium on Personal Indoor and
  Mobile Radio Communications (PIMRC)}, September 2011, pp. 614 --618.

\bibitem{Tang2010}
Q.~Hu and Z.~Tang, ``An improved power control strategy for cognitive radio
  networks with imperfect channel estimation,'' in \emph{Proc. 2010 6th
  International Conference on Wireless Communications Networking and Mobile
  Computing (WiCOM)}, September 2010, pp. 1--4.

\bibitem{Popovski2011}
R.~D. Taranto and P.~Popovski, ``Outage performance in cognitive radio systems
  with opportunistic interference cancelation,'' \emph{IEEE Trans. Wireless
  Commun.}, vol.~10, pp. 1280--1288, April 2011.

\bibitem{Pei2011}
Y.~Pei, Y.-C. Liang, K.~C. Teh, and K.~H. Li, ``Secure communication in
  multiantenna cognitive radio networks with imperfect channel state
  information,'' \emph{IEEE Trans. Signal Process.}, vol.~59, pp. 1683--1693,
  April 2011.

\bibitem{AlouiniISWCS2010}
Z.~Rezki and M.-S. Alouini, ``On the capacity of cognitive radio under limited
  channel state information,'' in \emph{Proc. 2010 7th International Symposium
  on Wireless Communication Systems (ISWCS)}, September 2010, pp. 1066 --1070.

\bibitem{Dey2012}
Y.-Y. He and S.~Dey, ``Throughput maximization in cognitive radio under peak
  interference constraints with limited feedback,'' \emph{IEEE Transactions on
  Vehicular Technology}, vol.~61, no.~3, pp. 1287 --1305, March 2012.

\bibitem{Duan2010}
R.~Duan, R.~Jantti, M.~Elmusrati, and R.~Virrankoski, ``Capacity for spectrum
  sharing cognitive radios with mrc diversity and imperfect channel information
  from primary user,'' in \emph{Proc. 2010 IEEE Global Telecommunications
  Conference (GLOBECOM 2010)}, December 2010, pp. 1 --5.

\bibitem{Gradshteyn2007}
I.~S. Gradshteyn and I.~M. Ryzhik, \emph{Table of Integrals, Series and
  Products}, 7th~ed.\hskip 1em plus 0.5em minus 0.4em\relax San Diego, CA:
  Academic Press, 2007.

\bibitem{Dmochowski2010}
P.~A. Dmochowski, H.~A. Suraweera, P.~J. Smith, and M.~Shafi, ``Impact of
  channel knowledge on cognitive radio system capacity,'' in \emph{Proc. 2010
  IEEE 72nd Vehicular Technology Conference Fall (VTC 2010-Fall)}, September
  2010, pp. 1--5.

\bibitem{Ahn2009}
K.~S. Ahn and R.~W. {Heath Jr.}, ``Performance analysis of maximum ratio
  combining with imperfect channel estimation in the presence of cochannel
  interferences,'' \emph{IEEE Trans. Wireless Commun.}, vol.~8, pp. 1080--1085,
  March 2009.

\bibitem{Kotz1994}
N.~Johnson, S.~Kotz, and N.~Balakrishnan, \emph{Continuous Univariate
  Distributions}, 2nd~ed.\hskip 1em plus 0.5em minus 0.4em\relax
  Wiley-Interscience, 1994, vol.~1.

\end{thebibliography}

\end{document}